\documentclass[aps,superscriptaddress,notitlepage,twocolumn]{revtex4-1}

\usepackage{graphicx}
\usepackage{amsmath, amssymb}
\usepackage{hyperref}
\usepackage[x11names, rgb, html]{xcolor}
\definecolor{sbase03}{HTML}{002B36}
\definecolor{sbase02}{HTML}{073642}
\definecolor{sbase01}{HTML}{586E75}
\definecolor{sbase00}{HTML}{657B83}
\definecolor{sbase0}{HTML}{839496}
\definecolor{sbase1}{HTML}{93A1A1}
\definecolor{sbase2}{HTML}{EEE8D5}
\definecolor{sbase3}{HTML}{FDF6E3}
\definecolor{syellow}{HTML}{B58900}
\definecolor{sorange}{HTML}{CB4B16}
\definecolor{sred}{HTML}{DC322F}
\definecolor{smagenta}{HTML}{D33682}
\definecolor{sviolet}{HTML}{6C71C4}
\definecolor{sblue}{HTML}{268BD2}
\definecolor{scyan}{HTML}{2AA198}
\definecolor{sgreen}{HTML}{859900}
\hypersetup{
  colorlinks = true,
  citecolor = {sred},
  urlcolor = {sblue}
}

\newcommand{\pd}[2]{\frac{\partial #1}{\partial #2}} 
 
\newcommand{\avg}[1]{\langle #1 \rangle}

\begin{document}
\title{Robust nonequilibrium pathways to microcompartment assembly}

\author{Grant M. Rotskoff}
\affiliation{Courant Institute of Mathematical Sciences, New York University, New York, NY 10002}
\affiliation{Department of Chemistry, University of California, Berkeley, CA 94720}
\author{Phillip L. Geissler} 
\affiliation{Department of Chemistry, University of California, Berkeley, CA 94720}

\date{\today}

\begin{abstract}
  Cyanobacteria sequester photosynthetic enzymes into microcompartments which facilitate the conversion of carbon dioxide into sugars. 
  Geometric similarities between these structures and self-assembling viral capsids have inspired models that posit microcompartments as stable equilibrium arrangements of the constituent proteins. 
  Here we describe a different mechanism for microcompartment assembly, one that is fundamentally nonequilibrium and yet highly reliable. 
  This pathway is revealed by simulations of a molecular model resolving the size and shape of a cargo droplet, and the extent and topography of an elastic shell. 
  The resulting metastable microcompartment structures closely resemble those of carboxysomes, with a narrow size distribution and faceted shells. 
  The essence of their assembly dynamics can be understood from a simpler mathematical model that combines elements of classical nucleation theory with continuum elasticity. 
  These results highlight important control variables for achieving nanoscale encapsulation in general, and for modulating the size and shape of carboxysomes in particular.
\end{abstract}

\maketitle

Spatial segregation is an ubiquitous strategy in biology for organizing the crowded, active viscera of the cell~\cite{Menon:2008bf,Kerfeld:2010vt,Hinzpeter:2017bw,Polka:2016fp}.
Viral capsids exemplify this organization at very small scales, sequestering genetic material from the cytosol and recapturing it for delivery to new hosts.  
Extensive work has explored the structure, stability, and assembly dynamics of viruses, highlighting generic design principles and physical origins of the spontaneous assembly process~\cite{Mateu:2013wd,Grime:2016ic,JasonDPerlmutter:2015hw,Rapaport:2008kb,Zandi:2004ev,Zlotnick:1994jb}.  
Overall capsid structure typically follows from the arrangements of neighboring proteins that are preferred by their noncovalent interactions.  
Strong preferences yield regular and highly stable structures, but at the same time impair kinetic accessibility by producing deep kinetic traps~\cite{Whitelam:2015cw}.

Bacterial microcompartments serve a very different biomolecular purpose from viruses but have striking structural similarities, namely, a quasi-icosahedral protein shell that assembles around a fluctuating cargo~\cite{Cameron:2013wa,Tanaka:2008wp,Kerfeld:2010vt,Lassila:2014co}.
This comparison raises the question: Do the same assembly principles, based on a balance between equilibrium stability and kinetic accessibility, apply to microcompartments as well?
Here we focus on a paradigmatic example of a microcompartment, the carboxysome.
Carboxysomes play an essential role in the carbon fixation pathway of photosynthetic cyanobacteria~\cite{Savage:2010ut}. 
The shell proteins of the carboxysome, which have hexameric and pentameric crystallographic structures, assemble to encapsulate a condensed globule of protein including the enzymes RuBisCO and carbonic anhydrase~\cite{Yeates:2007cp,Bobik:2015we,Yeates:2010kx}.  
These vital, organelle-like structures regulate a microscopic environment to enhance catalytic efficiency, which has made them an attractive target for bioengineering applications~\cite{Cai:2015bf,Frey:2016cn,Giessen:2017ia,Chen:2006kj}.

\begin{figure*}
  \begin{center}
  \includegraphics[width=0.65\linewidth]{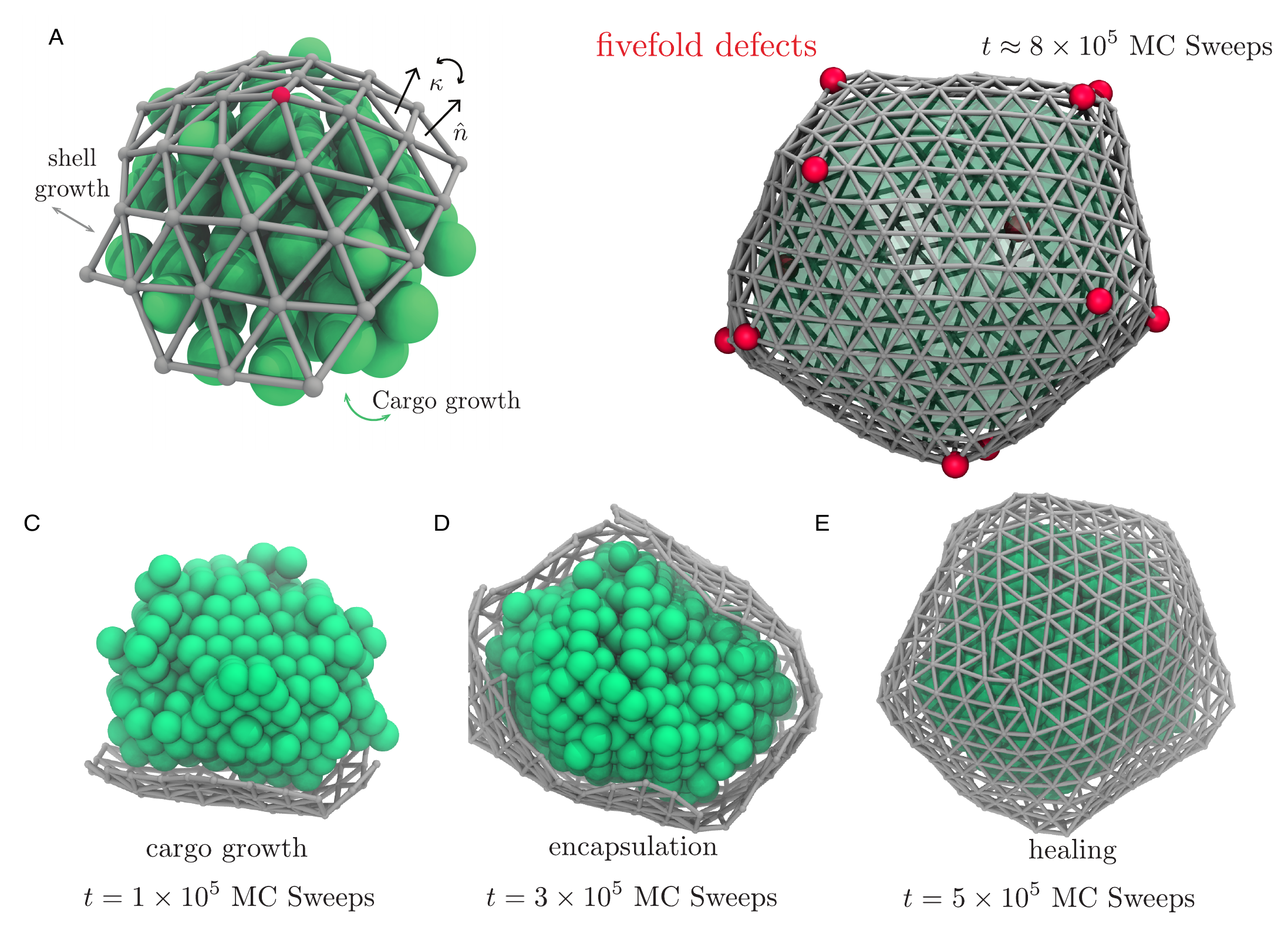}
\end{center}
  \caption{
    Model dynamics of microcompartment assembly. (A) Cargo monomers
    (green spheres) in our molecular model occupy sites of an FCC
    lattice, experiencing short-ranged attraction to their nearest
    neighbors and also to protein monomers comprising the shell.  Each
    shell monomer corresponds to a triangle in a discretized shell
    (gray) that resists bending and stretching according
    to an elastic Hamiltonian. Closure of the shell requires the presence of
    topological defects in the triangulated surface, vertices that are
    connected to only five neighbors. Red spheres here and in other
    figures highlight the locations of these defects.  (B) A fully
    assembled microcompartment includes at least twelve of these
    defects. Here and in other figures, we show the boundary of the
    cargo droplet as a translucent green surface. (C) At early time
    $t$ in the assembly process, high curvature of the cargo droplet
    limits shell growth to an area that is relatively flat while
    maintaining contact with the cargo. (D) Droplet growth reduces
    curvature until encapsulation becomes thermodynamically favorable
    and kinetically facile (Movie S1). (E) The nearly closed shell
    effectively halts cargo aggregation, but the approach to a
    simply connected envelope proceeds slowly as defects reposition
    and combine to heal grain boundaries.
  }
  \label{fig:traj}
\end{figure*}

Like some viral capsids, carboxysomes feature pronounced facets and vertices, and crystal structures of shell components suggest highly symmetric local protein arrangements~\cite{Yeates:2007cp,Yeates:2010kx,Sutter:2017ew}.  
Unlike in viral capsids, however, these apparent preferences do not directly indicate the assembled structure, nor do they clearly point to a characteristic microcompartment size~\cite{Cai:2009et}.  
Indeed, experimental evidence has not constrained a particular mechanism for assembly: live cell images support the notion that the cargo assembles first and is subsequently coated by the shell~\cite{Cameron:2013wa}, while direct observations of partially formed carboxysomes in electron micrographs point to a cooperative mechanism for assembly~\cite{Iancu:2010eb}.
Identifying essential variables for controlling or emulating carboxysome assembly awaits a clearer understanding of its dynamical pathways and underlying driving forces.

Theoretical and computational models for nanoshell assembly have
primarily followed approaches inspired by viral capsid assembly.
These approaches have emphasized the role of preferred shell curvature and
shell-cargo interactions as a template for the final
structure~\cite{Perlmutter:2016hz}.
For example, irreversible growth of shells comprising monomers
that prefer a bent binding geometry has been shown in simulations
to successfully assemble empty shells
~\cite{Wagner:2015jy,Hicks:2006gc}.
The assumption of such a preferred local curvature, however, is at odds
with structural models of the carboxysome based on crystallography of
shell proteins, which feature tiling of the shell by hexameric proteins
that appear to bind optimally with zero curvature~\cite{Yeates:2010kx,Sutter:2016,Iancu:2010eb}
Indeed, a number of experimental measurements bolster the view that
bacterial microcompartments are bounded by protein sheets that are
essentially flat away from localized regions of high curvature.  The
predominant constituents of diverse microcompartments (e.g.,
$\beta$-carboxysomes, $\alpha$-carboxsyomes, \emph{eut}
microcompartments, and \emph{pdu} microcompartments) crystallize as
layers of flat sheets comprising hexamers whose lateral contacts are
genetically conserved
~\cite{Yeates:2010kx}.  Recent atomic force
microscopy measurements have demonstrated
that these constituents also form flat monolayers in physiological buffer
conditions~\cite{Sutter:2016}.
Furthermore, \emph{in vivo} tomography
data strongly suggest that biological carboxysomes have extended flat
faces with curvature sharply localized at the joint of a
facet~\cite{Iancu:2010eb}.

Here we introduce and explore a model based on a mechanistic
perspective that is fundamentally different from the one commonly
applied to virus assembly and that does not require assuming an
innately preferred curvature for contacts between shell proteins.  The
basic components are a cargo species that is prone to aggregation, a
shell species that spontaneously forms flat,
hexagonally symmetric
elastic sheets, and an
attractive interaction between the inside of the shell and the cargo,
depicted in Figs. 1A, S1-S3.  These ingredients
appear to be
the
essential
constituents for carboxysome biogenesis \emph{in vitro}---mutagenesis
experiments have shown that pentameric proteins
sometimes presumed to stabilize shell curvature are in fact
not necessary for the formation of faceted shell
structures~\cite{Cai:2009et}.
For such a basic model, thermodynamic
considerations imply that finite encapsulated structures
have negligible weight at equilibrium (Sec. S3).  Nevertheless,
we show that regularly sized microcompartments can emerge reliably in
the course of natural dynamics.

\section{Methods}

We regard the protein shell of a growing carboxysome as
a thin elastic sheet, whose energy 
can be
discretized and expressed as a sum over the edges $ij$ of a
triangulated surface~\cite{Seung:1988ix}.
Its Hamiltonian is
\begin{equation}
  \mathcal{H}_{\rm shell} = \frac{\epsilon}{2} \sum_{ij} (l_{ij}-l_0)^2 + \frac{\kappa}{2} \sum_{ij} 1-\cos(\theta_{ij}) + \mathcal{H}_{\rm steric},
  \label{eq:Hdiscrete}
\end{equation}
where $l_{ij}$ is the length of an edge, whose deviations from the
preferred length $l_0$ experience a restoring force with stiffness
$\epsilon$. The bending rigidity $\kappa$ sets the energy scale for
developing a nonzero angle $\theta_{ij}$ between normal vectors of
triangles sharing the edge $ij$~\cite{Seung:1988ix}.

We associate each triangular face in the discretized sheet with a
protein monomer.  The term $\mathcal{H}_{\rm steric}$ in~\eqref{eq:Hdiscrete}
imposes steric constraints that prevent overlap of these monomers, by
placing a hard sphere of diameter $0.45 l_0$  at the center of each triangle
in the sheet.  
Because the deformations of individual shell proteins
are expected to be small relative to the bending fluctuations between
adjacent monomers, we work in a limit where $\epsilon l_0^2 \gg
\kappa.$ The cohesive interactions that bind adjacent shell proteins
in the sheet are accounted for separately, with an energy
$\mathcal{H}_{\rm bind}$ that tightly associates the vertices of contacting
monomers. In the limit that bound vertices coincide exactly, the
thermodynamic influence of $\mathcal{H}_{\rm bind}$ depends only on
the connectivity of the sheet, contributing a binding affinity factor
$K$ for each constrained vertex, as described in SI.

We represent aggregating cargo as an Ising lattice gas on an FCC lattice
with chemical potential $\mu_\textrm{c}$, restricted to configurations
with a single connected cargo droplet.
Nearest neighbors are coupled through an interaction energy $\epsilon_\textrm{c}$,
\begin{equation}
  \mathcal{H}_{\rm cargo} = -\mu_\textrm{c} \sum_{i=1}^{N_\textrm{L}} \sigma_i - \epsilon_\textrm{c} \sum_{\avg{ij}} \sigma_i \sigma_j,
\end{equation}
where $\sigma_i = \{0, 1\}$ for occupied and unoccupied lattice sites,
respectively, $\avg{ij}$ indicates that the sum is taken over
nearest neighbors, and $N_\textrm{L}$ is the number of lattice sites.  
The lattice spacing used throughout is $l_0,$ the
average length of a shell monomer edge.
The primary cargo species in a carboxysome, the protein RuBisCO, has a
diameter $L_0$ that is larger than $l_0$; adopting a finer lattice
spacing amounts to averaging approximately over configurational
fluctuations of RuBisCO about a close-packed lattice.

Interactions between cargo and shell species in our model mimic a
short-ranged directional attraction suggested by structural
data~\cite{Kinney:2011ps}, and the steric repulsion intrinsic to
compact macromolecules. For a particular lattice site $i$ and shell
monomer $j$, 
\begin{equation}
\mathcal{H}_\textrm{int}(\sigma_i, \hat{n}_j) = 
\begin{cases}
  -\gamma_\textrm{in}  &\textrm{if } -\hat{n}_j \in \mathcal{V}(i) \textrm{ and } \sigma_i = 1 \\
    +\gamma_\textrm{out} & \textrm{if } \hat{n}_j \in \mathcal{V}(i) \textrm{ and } \sigma_i = 1 \\
    0 & \textrm{otherwise},
\end{cases}
\end{equation}
where $\hat{n}_j$
is
the outward facing normal vector of shell monomer $j$.  
The notation $\in \mathcal{V}(i)$ means that the point specified by the vector lies within the volume $\mathcal{V}(i)$ occupied by lattice cell $i.$
The full interaction is the sum of $H_{\textrm{int}}$ over all $i$ and $j$.  
This interaction, which contributes a favorable energy if the 
inward-facing normal lies within a cargo-occupied lattice site
and an unfavorable energy if the outward-facing normal vector
does so, is depicted schematically in Fig. S5.

The thermal relaxation of a shell with a fixed number of vertices and
fixed connectivity, subject to the energetics described above,
can be simulated with a straightforward Monte Carlo
dynamics.  In each move, a single vertex
is chosen
at random.  As depicted in Fig. S2 (A), a random perturbation in
three-dimensional space is made to the selected vertex,
attempting to change its position from $a$ to $a'.$
The resulting energy difference,
$\Delta \mathcal{H}_{\rm shell} + \Delta \mathcal{H}_{\rm int}$,
is used in a standard Metropolis acceptance
criterion,
\begin{equation}
  \textrm{acc}(a\to a') = \min\left[ 1, e^{ -\beta
      (\Delta \mathcal{H}_{\rm shell} + \Delta \mathcal{H}_{\rm int})}
       \right].
\end{equation}
This procedure ensures that the configurations sampled in the Markov
chain are consistent with a Boltzmann distribution.

A Monte Carlo dynamics for growth of the shell is significantly more
involved.  In order that the procedure satisfy detailed balance, care
must be taken
to ensure that every step is reversible and
that algorithmic asymmetries of forward and
reverse moves are fully accounted for.
Our goal is to draw samples from a grand
canonical distribution,
\begin{equation}
  p(X, N_{\textrm{s}}, \{ \sigma \}) = \Xi^{-1} z_{\textrm{s}}^{N_{\textrm{s}}} e^{-\beta (\mathcal{H}_{\rm shell}
    +\mathcal{H}_{\rm bind} + \mathcal{H}_{\rm cargo} + \mathcal{H}_{\rm int})},
\end{equation}
where $\Xi$ denotes the grand canonical partition function, $X$ gives
the coordinates of the shell protein configuration, $N_{\textrm{s}}$ is the number
of shell monomers, and $z_{\textrm{s}}\propto e^{\beta \mu_{\textrm{s}}}$ is an activity set by
the concentration of free shell monomers in solution.
We implement two types of Monte Carlo moves
that allow the structure to grow,
increasing the number of monomers by adding new vertices at the edge
of the shell.  Fusion of pre-existing vertices can also occur,
representing the binding between monomers in the shell that are nearby
in space but not necessarily in connectivy.  Details of these moves
and their reverse counterparts, along with acceptance criteria that
ensure detailed balance with respect to $p(X, N_{\textrm{s}}, \{ \sigma\})$, are described in
Sec.~SI 1.  In microcompartment growth simulations, they are
performed in tandem with standard trial moves that change the
occupation state of the cargo lattice, acting to grow or shrink the
cargo droplet.

The routes of Monte Carlo trajectories propagated in this way are
influenced by energetic parameters like $\epsilon$, $\kappa$, and
$\mu$, and by the shell binding affinity $K$.
The pathways are also shaped by the relative
frequencies of proposing each type of move. A basic time scale $\tau$
is set by the duration of a ``sweep'' in which each vertex experiences
a single attempted spatial displacement (on average). For every $n_\textrm{c}$
sweeps we attempt roughly $N_{\rm boundary}$ moves that add or remove
material at the surface of the cargo droplet, where $N_{\rm boundary}$
is the number of lattice sites defining the droplet surface. This procedure
establishes an effective rate $k_\textrm{c}^0 = (2 n_\textrm{c} \tau)^{-1}$ at which
cargo monomers arrive from solution at a given surface
site. Similarly, a basic rate $k_{\textrm{s}}^0 = (2 n_{\textrm{s}} \tau)^{-1}$ for shell
monomer arrival at a given site on the shell's perimeter is
established by attempting $N_{\textrm{perim}}$ shell monomer
addition/removal moves for every $n_{\textrm{s}}$ sweeps, where $N_{\textrm{perim}}$
is the number of edges at the shell boundary. In experiments these
arrival rates are approximately proportional to the solution
concentrations of the respective monomers. 
As a result,
microcompartment mass changes slowly on the time scale of elastic
fluctuations.  Dynamics of vertex fusion and fission also proceed much more
rapidly than growth.
In essence, all
processes other than monomer addition and removal follow along
adiabatically.
For the fate of assembly, the key dynamical parameter in our model is
therefore the ratio $k_\textrm{c}^0/k_{\textrm{s}}^0$ of arrival rates for cargo and shell
species.

\section{Molecular simulations}

Fig.~\ref{fig:traj} depicts an example assembly pathway of this molecular model.  
Trajectories are initiated with a small droplet of cargo
(comprising a few hundred cargo monomers)
and a handful of proximate shell monomers, as described in Sec. S5.  
Under conditions favorable for assembly,
such a droplet faces no thermodynamic barrier to growth; absent interactions with the shell, its radius $R$ would increase at a constant average rate as cargo material arrives from the droplet's surroundings.  
Growth of the shell, on the other hand, is impeded by the energetic cost of wrapping an elastic sheet around a highly curved object.  
In early stages of this trajectory, shown in Fig.~\ref{fig:traj} (C), the net cost of encapsulation is considerable, and the population of shell monomers remains small as a result.

As the droplet grows,
this elastic penalty
is eventually overwhelmed by
attractions between shell and cargo, similar to the mechanism of
curvature generation by nanoparticles adsorbed on membrane
surfaces~\cite{Bahrami:2012gb, vanderWel:2016du, Saric:2012hb}.  At a
characteristic droplet size $R^*$, encapsulation becomes
thermodynamically favorable.  If the arrival rate of shell monomers is
much higher than that of cargo (e.g., due to a higher concentration in
solution), then a nearly complete shell will quickly develop,
Fig.~\ref{fig:traj} (D), hampering the incorporation of additional
cargo.  The nascent shell that results, while sufficient to block
cargo arrival, is highly defective and far from closed.  A slow
healing process ensues, dominated by relaxation of grain boundaries
between growth faces, Fig.~\ref{fig:traj} (E).  This annealing process
is, in part, achieved by relocating topological defects in the
structure, a phenomenon that has been previously encountered in ground
state calculations for closed elastic
shells~\cite{Funkhouser:2012hea}.

In the vast majority of the hundreds of assembly trajectories we have
generated, healing leads ultimately to a completely closed structure
with exactly 12 five-fold defects.  Placement of these defects is
often irregular and unlikely to yield minimum elastic energy, but
further evolution of the structure is extraordinarily slow.
Subsequent defect dynamics could produce more ideal shell structures
as in Ref.~\cite{Funkhouser:2012hea}, and transient shell opening
could allow additional cargo growth. But these relaxation processes
require the removal of shell monomers that are bound to several others
and that interact strongly with enclosed cargo. Under the conditions
of interest, the time scale for such a removal is vastly longer than
the assembly trajectories we propagate. 
Our model microcompartments are
equilibrated with respect to neither shell geometry nor droplet size,
yet they are profoundly metastable, requiring extremely rare events to
advance towards the true equilibrium state.

In addition to generating molecular trajectories of microcompartment
assembly, we have performed umbrella sampling simulations to compute
the equilibrium free energy of the molecular model as a function of
$N_\textrm{s}$ and the number $N_\textrm{c}$ of cargo
monomers. Results, presented in Sec. S4, underscore the mechanistic
features described above: a small critical nucleus size for cargo
condensation, and a size-dependent thermodynamic bias on encapsulation
that becomes sharply favorable at a characteristic length scale.  In
the next section we examine the origins and consequences of these
features in a more idealized context.

\begin{figure}

  \includegraphics[width=\linewidth]{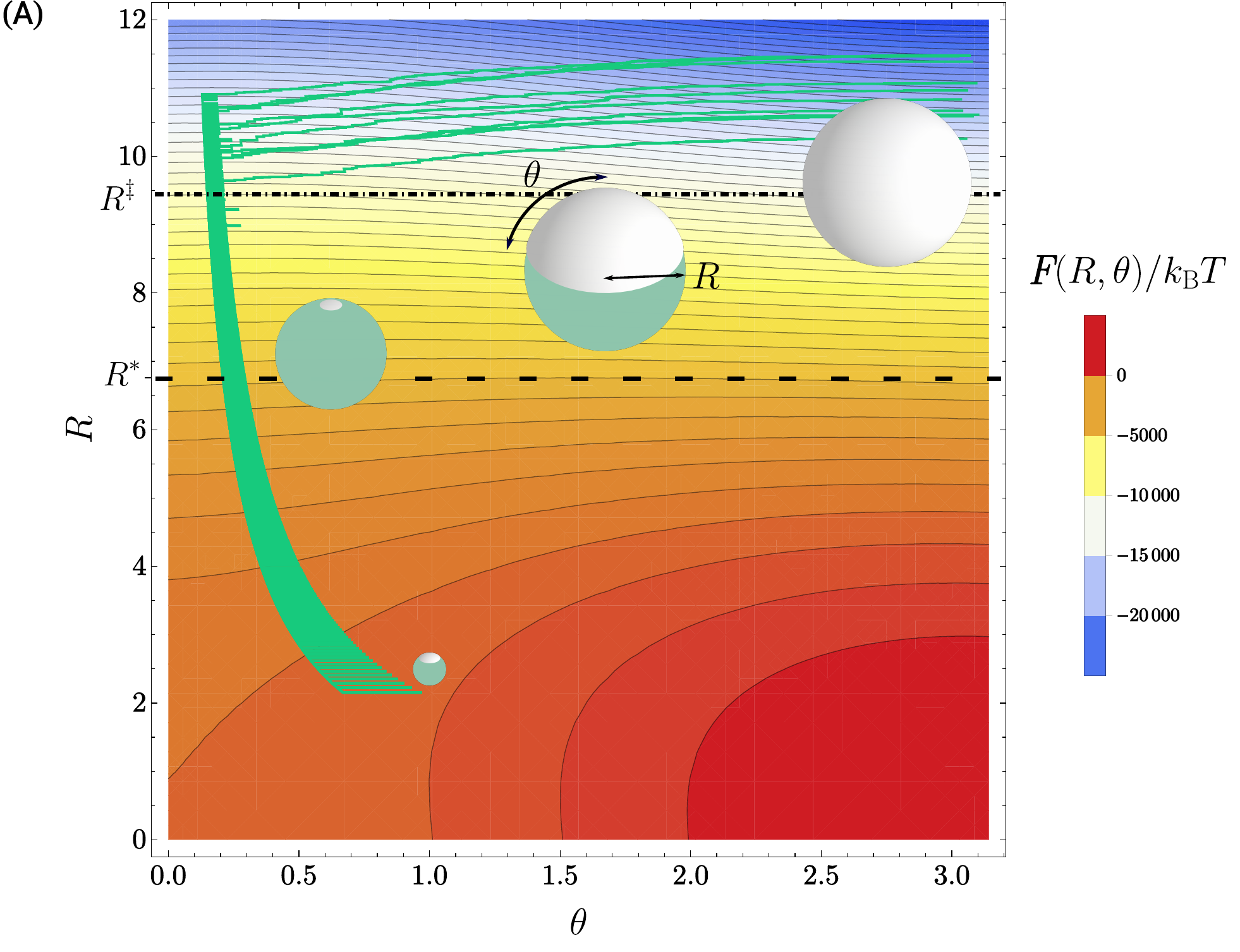}

  \qquad\includegraphics[width=\linewidth]{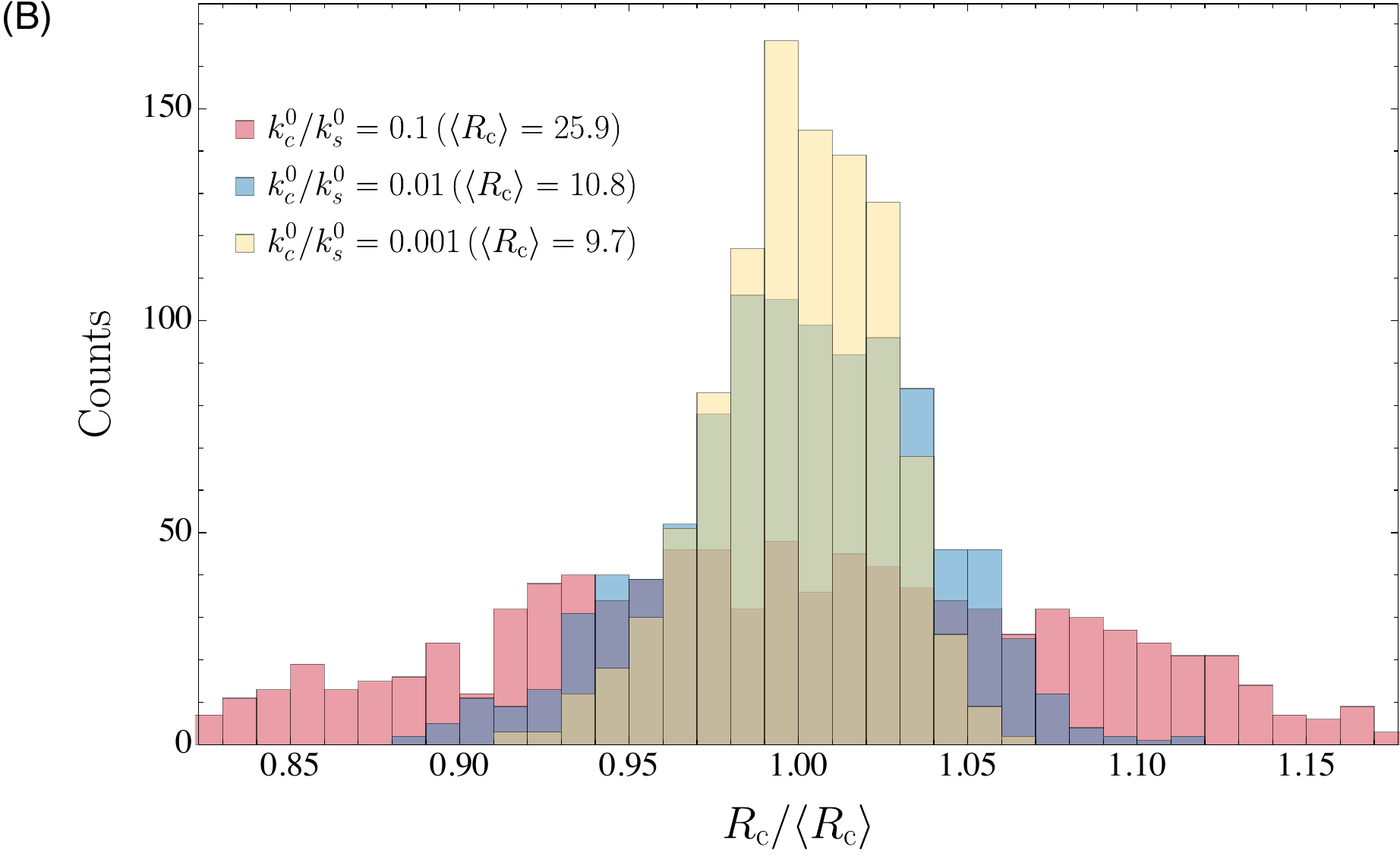}
  \caption{
    Minimalist model for dynamics of cargo growth and encapsulation.  (A)
    Our phenomenological model resolves only the radius $R$ of a cargo
    droplet, and the polar angle $\theta$ of a spherical cap that
    coats it. Black contours indicate lines of constant free energy
    $F(R,\theta)$. Green lines show the course of ten
     kinetic Monte Carlo trajectories under conditions
    favorable for microcompartment assembly 
    ($k_{\textrm{c}}^0/k_{\textrm{s}}^0 = 0.001$, all other parameters given in Sec. S4). Several structures along the
    assembly trajectory are shown near the corresponding values of $R$
    and $\theta$.  Encapsulation is thermodynamically favorable for
    $R>R^*$ (lower dashed line). The free energy barrier to
    encapsulation is smaller than the thermal energy $k_{\rm B}T$ for
    $R>R^\ddagger$ (Eq. S63)
    (dot-dashed line). (B) Histograms of microcompartment radius
    when $\theta$ reaches $\pi$ (at which point dynamics cease).  Ten
    thousand independent trajectories were collected for $k_{\textrm{c}}^0/k_{\textrm{s}}^0$
    = 0.1, 0.01, and 0.001.
    Radius values are given in a unit of length $\ell$ that is comparable to the shell monomer size (see Eq. S37).
  }
  \label{fig:fes}
\end{figure}

\section{Minimalist model}
The scenario described above for our molecular model can be cast in a
simpler light by considering as dynamical variables only the amounts
of cargo and shell material in an idealized geometry.  In this
minimalist approach we focus on the radius $R$ of a spherical droplet
and the polar angle $\theta$ subtended by a contacting spherical cap
of shell material, as depicted in Fig.~\ref{fig:fes} (A). Taking the
cargo and shell species to have uniform densities $\rho$ and $\nu$
within the droplet and cap, respectively, the geometric parameters $R$
and $\theta$ can be simply related to the monomer populations
$N_\textrm{c} = (4\pi\rho/3)R^3$ and $N_\textrm{s} = 2\pi \nu R^2
(1-\cos\theta)$.

In the
spirit of classical nucleation theory, we estimate a free energy
landscape in the space of $R$ and $\theta$ through considerations of
surface tension and bulk thermodynamics.  For our system these
contributions include free energy of the condensed cargo phase,
surface tension of a bare cargo droplet, free energy of a macroscopic
elastic shell, and line tension of a finite shell, together with the
energetics of shell bending and shell-cargo attraction.  As shown in
the Sec. S3, this free energy $F(R,\theta)$ can be written in the form

\begin{widetext}
\begin{equation}
  F(R, \theta) = a_\textrm{c} \left[ \left( \frac{R}{R_\textrm{c}} \right)^2 - \left( \frac{R}{R_\textrm{c}} \right)^3 \right] + a_\textrm{s} \left[ \left\{ \left(\frac{R^*}{R_\textrm{s}} \right)^2 - \left(\frac{R}{R_\textrm{s}} \right)^2 \right\}\left(1-\cos \theta\right) + \frac{R}{R_\textrm{s}} \sin \theta \right],
  \label{eq:fes}
\end{equation}
\end{widetext}

\noindent
where the energy scales $a_\textrm{c}$ and $a_\textrm{s}$ and the length scales $R_\textrm{c}$, $R^*$, and $R_\textrm{s}$ (all of which can be related to parameters of the molecular model) transparently reflect the thermodynamic biases governing growth and encapsulation.
The function $F(R,\theta)$ can be minimum only at $R=0$ and $R=\infty$.  As in the molecular model, finite microcompartments can appear and persist only as kinetically trapped nonequilibrium states.

Contours of this schematic free energy surface, plotted in
Fig.~\ref{fig:fes} (A), are consistent with basic thermodynamic trends
observed for the molecular model. Most importantly, they manifest the
prohibitive cost of encapsulating small droplets.  Only for radii
$R>R^*$ is shell growth thermodynamically favorable.  This
characteristic length scale is determined by a straightforward balance
among the shell's bending rigidity, the macroscopic driving force for
flat shell growth, and the attraction between shell and cargo, as
detailed in Sec. S3.  Whether encapsulation occurs immediately once
the droplet exceeds the critical size $R^*$ depends on kinetic factors
that are only partly governed by $F(R,\theta)$.  Line tension of the
shell, represented by the final term in Eq.~\ref{eq:fes}, effects a
barrier to encapsulation.  As the droplet size increases, the height
of this barrier declines, and encapsulation becomes facile only when
it is sufficiently low. As with our molecular model, the subsequent
dynamics may deviate significantly from equilibrium expecations based
on the free energy surface. To explore these nonequilibrium outcomes,
we must supplement the free energy surface $F(R,\theta)$ with a
correspondingly minimal set of dynamical rules.

\begin{figure*}
\begin{center}
  \includegraphics[width=\linewidth]{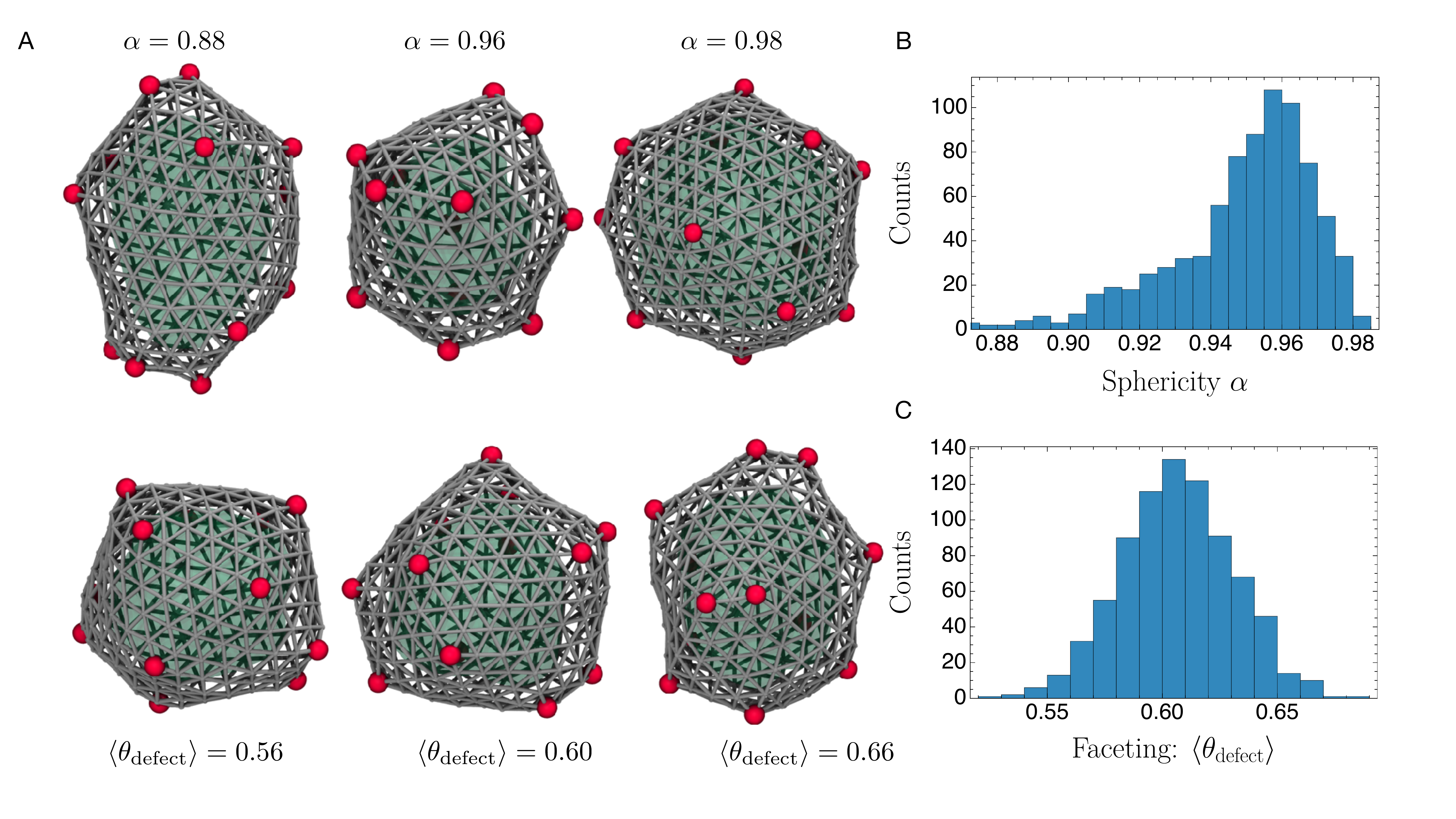}
  \end{center}
  \caption{
    Structural variation among microcompartments generated by our
    molecular model. (A) Structures from the ensemble evincing the distinct regimes of sphericity and faceting. Note that two dimensional projections can obscure some of the variation. (B) Typical
    assemblies are not substantially elongated, as indicated by a
    histogram of the sphericity $\alpha$ (Sec. S7). (C) Faceting
    necessitates relatively sharp angles between shell monomers
    surrounding a fivefold defect (Sec. S7). A histogram of this angle
    $\avg{\theta_\textrm{defect}}$ (averaged over all defects in
    an assembled structure) underscores the typically strong faceting
    visible in (A).
  }
  \label{fig:dists}
\end{figure*}

We take the rate of shell monomer addition to be a product of the
per-edge rate of monomer arrival $k_{\textrm{s}}^0$ and the number of
edges that can bind new monomers,
\begin{equation}
  k_{\textrm{s}}^+(N_{\textrm{s}}, N_{\textrm{c}})   = k_{\textrm{s}}^0 \frac{2\pi
    R}{l_0} \sin \theta
  \label{eq:nsconv}
\end{equation}
Similarly accounting for the number of sites
where cargo can be added, we take the rate of cargo monomer
addition to be
\begin{equation}
  k_{\textrm{c}}^+(N_{\textrm{s}}, N_{\textrm{c}}) = k_{\textrm{c}}^0 4\pi \rho L_0 R^2 (1+\cos\theta),
    \label{eq:ncconv}
\end{equation}
The ratio of addition and removal rates is completely specified by the constraint of detailed balance.
Defining the shell monomer and cargo monomer free energy differences
\begin{eqnarray}
\Delta F_{\textrm{s}} &= F(N_{\textrm{s}}, N_{\textrm{c}}) - F(N_{\textrm{s}}-1, N_{\textrm{c}}), \\
\Delta F_{\textrm{c}} &= F(N_{\textrm{s}}, N_{\textrm{c}}) - F(N_{\textrm{s}}, N_{\textrm{c}}-1), \\
\end{eqnarray}
we require that the rates of monomer removal satisfy
\begin{equation} 
k_{\textrm{s}}^-(N_{\textrm{s}}, N_{\textrm{c}}) = k_{\textrm{s}}^+(N_{\textrm{s}}-1,N_{\textrm{c}}) e^{ \beta \Delta F_{\textrm{s}} },
\label{eq:ksm}
\end{equation}
and
\begin{equation} 
k_{\textrm{c}}^-(N_{\textrm{s}}, N_{\textrm{c}}) = k_{\textrm{s}}^+(N_{\textrm{s}},N_{\textrm{c}}-1) e^{ \beta \Delta F_{\textrm{c}} }.
\label{eq:kcm}
\end{equation}

Stochastic trajectories for this minimalist description were generated
by advancing the populations $N_{\textrm{s}}$ and $N_{\textrm{c}}$ in time with a kinetic
Monte Carlo algorithm based on the transition rates $k_{\textrm{s}}^+$, $k_{\textrm{c}}^+$,
$k_{\textrm{s}}^-$, and $k_{\textrm{c}}^-$. By construction, these rates are consistent with
the equilibrium statistics implied by~\eqref{eq:fes}, but the geometry
dependence of
addition and removal rates
ensures that typical trajectories do not simply
follow a gradient descent on the free energy surface $F(R,\theta)$.
The sampled paths shown in Fig.~\ref{fig:fes} (A) (initiated with
$N_{\textrm{s}}=1$ and $N_{\textrm{c}}=10$, and terminated when $N_{\textrm{s}} \geq 2 N_{\textrm{c}}^{2/3}$ as
described in Sec. S4) indeed exhibit strong geometric biases.

Most notably, $\theta$ increases rapidly at late times in these
successful assembly trajectories, despite equilibrium forces that
encourage droplet growth at the expense of reduced shell coverage.
This route reflects an imbalance in the base addition rates
$k_{\textrm{s}}^0$ and $k_{\textrm{s}}^0$ of cargo and shell material,
which suppresses droplet growth while the shell develops.  If the
addition of shell material does not outpace that of cargo, then the
angle $\theta$ decreases in time---the microcompartment grows but
never advances towards closure, Figs. S5 and S6.  Since the addition
rate of cargo scales with the droplet's surface area, while the rate
of adding shell material scales only with the shell's perimeter,
successful encapsulation generally requires that the imbalance in
addition rates be substantial, which could be straightforwardly
achieved through a large excess of shell monomers in solution relative
to cargo.  When the shell is able to envelop the cargo with only
modest changes in droplet size, the distribution of sizes can be quite
narrow, as shown in Fig.~\ref{fig:fes} (B).

\section{Discussion}
The minimalist model defined by Eqs. 7-13 explains the general course
of our more detailed molecular assembly trajectories.  But its
assumption of ideally spherical geometry is a rough approximation, as
evidenced by asymmetric droplet shapes in the molecular trajectory of
Fig. 1.  Fluctuations in cargo droplet shape are discouraged by its
surface tension, but they are inevitable on small length scales.  In
extreme cases they lead to highly aspherical microcompartments akin to
distended structures that have been observed in
experiments~\cite{Cai:2009et}.  More common shape fluctuations have
mechanistic influences that are more subtle but nonetheless
important.  These act to reduce local curvature and thus facilitate
growth of a nearly flat shell patch over considerable areas even in
the early stages of assembly.  Enveloping the entire cargo globule
still awaits sufficient cargo growth to reduce the average curvature
below a threshold of roughly $1/R^*$.  The necessary curvature that
ultimately develops in the shell is also spatially heterogeneous.
Protein sheets, which generally resist stretching more strongly than
bending, tend to concentrate curvature at topological defects and
along lines connecting them~\cite{Bowick:2002cf}.  As a kinetic
consequence, bending is advanced primarily by discrete, localized
events, in particular the formation of fivefold defects during shell
growth.  Indeed, the dynamics of encapsulation are noticeably
punctuated by the appearance of topological defects, as seen in Movie
S1.

Systems out of equilibrium often support rich spatial and temporal patterns, but in many cases they are highly variable and challenging to harness or control.  
By contrast, the encapsulation process we have
described is quite reliable. 
Fig.~\ref{fig:dists} (A) shows closed structures resulting from assembly
trajectories of our molecular model.  
They are not perfectly icosahedral (Figs. S7 (A) and S7 (B)), echoing the discernibly nonideal geometries observed in electron micrographs of carboxysomes~\cite{Iancu:2010eb}.  
Our assemblies are nonetheless roughly isotropic and distinctly faceted, as quantified by the distributions of sphericity and defect angles shown in Fig.~\ref{fig:dists} (B) and Fig. S7.  
Gently curved regions do sometimes appear, as they do in the laboratory.
In our model this smoothness reflects the presence of paired defects (one fivefold and one sevenfold) that sometimes form in the process of shell closure, screening the elastic interactions that give rise to faceting~\cite{Bowick:2002cf}.

In published~\cite{Perlmutter:2016hz} and concurrent work, Hagan et
al. have computationally examined assembly dynamics of
microcompartments stabilized by spontaneous curvature of the protein
shell.  Experimental data suggest that key shell proteins of the
carboxysome tend, by contrast, to bind in an unbent geometry. But
existing measurements do not definitively exclude a role for
mechanically preferred curvature in carboxysome assembly. Quantitative
characterization of the elastic properties of shell protein sheets in
the absence of cargo would help clarify this issue.  Systematic
measurement of microcompartment geometry as a function of protein
concentrations would provide a more stringent test of our
nonequilibrium mechanistic hypothesis, which can be achieved with reconstitution experiments.

The nonequilibrium assembly mechanism revealed by our model has a
particularly spare set of essential requirements: One needs only a
monomer that assembles planar, elastic sheets and has a directional
affinity for cargo that is prone to aggregation.  Tuning the
concentrations of these components provides a direct route to
generating monodisperse, nanoscale, self-organizing structures.  The
simplicity of the process presents opportunities to investigate the
dynamics of shell assembly in highly controlled experimental settings,
for example using colloidal particles. Such a platform has already
been demonstrated for a related example of kinetically defined
microsctructure assembly, namely bicontinuous gels stabilized by
colloidal interfaces~\cite{Stratford:2005}.  Our results suggest that 
kinetic control can be used for precise, nanoscale assembly in a surprisingly general context.

\acknowledgements{
We thank Michael Hagan, Christoph Dellago, David Savage, Laura
Armstrong, Avi Flamholz, and Luke Oltrogge for helpful conversations.
GMR acknowledges support from National Science Foundation Graduate
Research Fellowship and the James S. McDonnell Foundation. 
PLG acknowledges support from the Erwin Schrödinger International
Institute for Mathematics and Physics. This work
was supported by the US Department of Energy, Office of Basic Energy
Sciences, through the Chemical Sciences Division (CSD) of the Lawrence
Berkeley National Laboratory (LBNL), under Contract
DE-AC02-05CH11231(PLG).
}

\bibliography{shells}

\begin{thebibliography}{40}%
\makeatletter
\providecommand \@ifxundefined [1]{%
 \@ifx{#1\undefined}
}%
\providecommand \@ifnum [1]{%
 \ifnum #1\expandafter \@firstoftwo
 \else \expandafter \@secondoftwo
 \fi
}%
\providecommand \@ifx [1]{%
 \ifx #1\expandafter \@firstoftwo
 \else \expandafter \@secondoftwo
 \fi
}%
\providecommand \natexlab [1]{#1}%
\providecommand \enquote  [1]{``#1''}%
\providecommand \bibnamefont  [1]{#1}%
\providecommand \bibfnamefont [1]{#1}%
\providecommand \citenamefont [1]{#1}%
\providecommand \href@noop [0]{\@secondoftwo}%
\providecommand \href [0]{\begingroup \@sanitize@url \@href}%
\providecommand \@href[1]{\@@startlink{#1}\@@href}%
\providecommand \@@href[1]{\endgroup#1\@@endlink}%
\providecommand \@sanitize@url [0]{\catcode `\\12\catcode `\$12\catcode
  `\&12\catcode `\#12\catcode `\^12\catcode `\_12\catcode `\%12\relax}%
\providecommand \@@startlink[1]{}%
\providecommand \@@endlink[0]{}%
\providecommand \url  [0]{\begingroup\@sanitize@url \@url }%
\providecommand \@url [1]{\endgroup\@href {#1}{\urlprefix }}%
\providecommand \urlprefix  [0]{URL }%
\providecommand \Eprint [0]{\href }%
\providecommand \doibase [0]{http://dx.doi.org/}%
\providecommand \selectlanguage [0]{\@gobble}%
\providecommand \bibinfo  [0]{\@secondoftwo}%
\providecommand \bibfield  [0]{\@secondoftwo}%
\providecommand \translation [1]{[#1]}%
\providecommand \BibitemOpen [0]{}%
\providecommand \bibitemStop [0]{}%
\providecommand \bibitemNoStop [0]{.\EOS\space}%
\providecommand \EOS [0]{\spacefactor3000\relax}%
\providecommand \BibitemShut  [1]{\csname bibitem#1\endcsname}%
\let\auto@bib@innerbib\@empty
\bibitem [{\citenamefont {Menon}\ \emph {et~al.}(2008)\citenamefont {Menon},
  \citenamefont {Dou}, \citenamefont {Heinhorst}, \citenamefont {Shively},\
  and\ \citenamefont {Cannon}}]{Menon:2008bf}%
  \BibitemOpen
  \bibfield  {author} {\bibinfo {author} {\bibfnamefont {B.~B.}\ \bibnamefont
  {Menon}}, \bibinfo {author} {\bibfnamefont {Z.}~\bibnamefont {Dou}}, \bibinfo
  {author} {\bibfnamefont {S.}~\bibnamefont {Heinhorst}}, \bibinfo {author}
  {\bibfnamefont {J.~M.}\ \bibnamefont {Shively}}, \ and\ \bibinfo {author}
  {\bibfnamefont {G.~C.}\ \bibnamefont {Cannon}},\ }\href {\doibase
  10.1371/journal.pone.0003570} {\bibfield  {journal} {\bibinfo  {journal}
  {PLOS ONE}\ }\textbf {\bibinfo {volume} {3}},\ \bibinfo {pages} {e3570}
  (\bibinfo {year} {2008})}\BibitemShut {NoStop}%
\bibitem [{\citenamefont {Kerfeld}\ \emph {et~al.}(2010)\citenamefont
  {Kerfeld}, \citenamefont {Heinhorst},\ and\ \citenamefont
  {Cannon}}]{Kerfeld:2010vt}%
  \BibitemOpen
  \bibfield  {author} {\bibinfo {author} {\bibfnamefont {C.~A.}\ \bibnamefont
  {Kerfeld}}, \bibinfo {author} {\bibfnamefont {S.}~\bibnamefont {Heinhorst}},
  \ and\ \bibinfo {author} {\bibfnamefont {G.~C.}\ \bibnamefont {Cannon}},\
  }\href {\doibase 10.1146/annurev.micro.112408.134211} {\bibfield  {journal}
  {\bibinfo  {journal} {Annu. Rev. Microbiol.}\ }\textbf {\bibinfo {volume}
  {64}},\ \bibinfo {pages} {391} (\bibinfo {year} {2010})}\BibitemShut
  {NoStop}%
\bibitem [{\citenamefont {Hinzpeter}\ \emph {et~al.}(2017)\citenamefont
  {Hinzpeter}, \citenamefont {Gerland},\ and\ \citenamefont
  {Tostevin}}]{Hinzpeter:2017bw}%
  \BibitemOpen
  \bibfield  {author} {\bibinfo {author} {\bibfnamefont {F.}~\bibnamefont
  {Hinzpeter}}, \bibinfo {author} {\bibfnamefont {U.}~\bibnamefont {Gerland}},
  \ and\ \bibinfo {author} {\bibfnamefont {F.}~\bibnamefont {Tostevin}},\
  }\href {\doibase 10.1016/j.bpj.2016.11.3194} {\bibfield  {journal} {\bibinfo
  {journal} {Biophys. J}\ }\textbf {\bibinfo {volume} {112}},\ \bibinfo {pages}
  {767} (\bibinfo {year} {2017})}\BibitemShut {NoStop}%
\bibitem [{\citenamefont {Polka}\ \emph {et~al.}(2016)\citenamefont {Polka},
  \citenamefont {Hays},\ and\ \citenamefont {Silver}}]{Polka:2016fp}%
  \BibitemOpen
  \bibfield  {author} {\bibinfo {author} {\bibfnamefont {J.~K.}\ \bibnamefont
  {Polka}}, \bibinfo {author} {\bibfnamefont {S.~G.}\ \bibnamefont {Hays}}, \
  and\ \bibinfo {author} {\bibfnamefont {P.~A.}\ \bibnamefont {Silver}},\
  }\href {\doibase 10.1101/cshperspect.a024018} {\bibfield  {journal} {\bibinfo
   {journal} {Cold Spring Harb. Perspect. Biol.}\ }\textbf {\bibinfo {volume}
  {8}},\ \bibinfo {pages} {a024018} (\bibinfo {year} {2016})}\BibitemShut
  {NoStop}%
\bibitem [{\citenamefont {Mateu}(2013)}]{Mateu:2013wd}%
  \BibitemOpen
  \bibfield  {author} {\bibinfo {author} {\bibfnamefont {M.~G.}\ \bibnamefont
  {Mateu}},\ }\href {\doibase http://dx.doi.org/10.1016/j.abb.2012.10.015}
  {\bibfield  {journal} {\bibinfo  {journal} {Arch. Biochem. Biophys.}\
  }\textbf {\bibinfo {volume} {531}},\ \bibinfo {pages} {65} (\bibinfo {year}
  {2013})}\BibitemShut {NoStop}%
\bibitem [{\citenamefont {Grime}\ \emph {et~al.}(2016)\citenamefont {Grime},
  \citenamefont {Dama}, \citenamefont {Ganser-Pornillos}, \citenamefont
  {Woodward}, \citenamefont {Jensen}, \citenamefont {Yeager},\ and\
  \citenamefont {Voth}}]{Grime:2016ic}%
  \BibitemOpen
  \bibfield  {author} {\bibinfo {author} {\bibfnamefont {J.~M.~A.}\
  \bibnamefont {Grime}}, \bibinfo {author} {\bibfnamefont {J.~F.}\ \bibnamefont
  {Dama}}, \bibinfo {author} {\bibfnamefont {B.~K.}\ \bibnamefont
  {Ganser-Pornillos}}, \bibinfo {author} {\bibfnamefont {C.~L.}\ \bibnamefont
  {Woodward}}, \bibinfo {author} {\bibfnamefont {G.~J.}\ \bibnamefont
  {Jensen}}, \bibinfo {author} {\bibfnamefont {M.}~\bibnamefont {Yeager}}, \
  and\ \bibinfo {author} {\bibfnamefont {G.~A.}\ \bibnamefont {Voth}},\ }\href
  {\doibase 10.1038/ncomms11568} {\bibfield  {journal} {\bibinfo  {journal}
  {Nat. Commun.}\ }\textbf {\bibinfo {volume} {7}},\ \bibinfo {pages} {11568}
  (\bibinfo {year} {2016})}\BibitemShut {NoStop}%
\bibitem [{\citenamefont {Perlmutter}\ and\ \citenamefont
  {Hagan}(2015)}]{JasonDPerlmutter:2015hw}%
  \BibitemOpen
  \bibfield  {author} {\bibinfo {author} {\bibfnamefont {J.~D.}\ \bibnamefont
  {Perlmutter}}\ and\ \bibinfo {author} {\bibfnamefont {M.~F.}\ \bibnamefont
  {Hagan}},\ }\href {\doibase 10.1146/annurev-physchem-040214-121637}
  {\bibfield  {journal} {\bibinfo  {journal} {Annu. Rev. Phys. Chem.}\ }\textbf
  {\bibinfo {volume} {66}},\ \bibinfo {pages} {217} (\bibinfo {year}
  {2015})}\BibitemShut {NoStop}%
\bibitem [{\citenamefont {Rapaport}(2008)}]{Rapaport:2008kb}%
  \BibitemOpen
  \bibfield  {author} {\bibinfo {author} {\bibfnamefont {D.~C.}\ \bibnamefont
  {Rapaport}},\ }\href {\doibase 10.1103/PhysRevLett.101.186101} {\bibfield
  {journal} {\bibinfo  {journal} {Phys. Rev. Lett.}\ }\textbf {\bibinfo
  {volume} {101}},\ \bibinfo {pages} {186101} (\bibinfo {year}
  {2008})}\BibitemShut {NoStop}%
\bibitem [{\citenamefont {Zandi}\ \emph {et~al.}(2004)\citenamefont {Zandi},
  \citenamefont {Reguera}, \citenamefont {Bruinsma}, \citenamefont {Gelbart},\
  and\ \citenamefont {Rudnick}}]{Zandi:2004ev}%
  \BibitemOpen
  \bibfield  {author} {\bibinfo {author} {\bibfnamefont {R.}~\bibnamefont
  {Zandi}}, \bibinfo {author} {\bibfnamefont {D.}~\bibnamefont {Reguera}},
  \bibinfo {author} {\bibfnamefont {R.~F.}\ \bibnamefont {Bruinsma}}, \bibinfo
  {author} {\bibfnamefont {W.~M.}\ \bibnamefont {Gelbart}}, \ and\ \bibinfo
  {author} {\bibfnamefont {J.}~\bibnamefont {Rudnick}},\ }\href {\doibase
  10.1073/pnas.0405844101} {\bibfield  {journal} {\bibinfo  {journal} {Proc.
  Natl. Acad. Sci. U.S.A.}\ }\textbf {\bibinfo {volume} {101}},\ \bibinfo
  {pages} {15556} (\bibinfo {year} {2004})}\BibitemShut {NoStop}%
\bibitem [{\citenamefont {Zlotnick}(1994)}]{Zlotnick:1994jb}%
  \BibitemOpen
  \bibfield  {author} {\bibinfo {author} {\bibfnamefont {A.}~\bibnamefont
  {Zlotnick}},\ }\href {\doibase http://dx.doi.org/10.1006/jmbi.1994.1473}
  {\bibfield  {journal} {\bibinfo  {journal} {J. Mol. Biol.}\ }\textbf
  {\bibinfo {volume} {241}},\ \bibinfo {pages} {59} (\bibinfo {year}
  {1994})}\BibitemShut {NoStop}%
\bibitem [{\citenamefont {Whitelam}\ and\ \citenamefont
  {Jack}(2015)}]{Whitelam:2015cw}%
  \BibitemOpen
  \bibfield  {author} {\bibinfo {author} {\bibfnamefont {S.}~\bibnamefont
  {Whitelam}}\ and\ \bibinfo {author} {\bibfnamefont {R.~L.}\ \bibnamefont
  {Jack}},\ }\href {\doibase 10.1146/annurev-physchem-040214-121215} {\bibfield
   {journal} {\bibinfo  {journal} {Annu. Rev. Phys. Chem.}\ }\textbf {\bibinfo
  {volume} {66}},\ \bibinfo {pages} {143} (\bibinfo {year} {2015})}\BibitemShut
  {NoStop}%
\bibitem [{\citenamefont {Cameron}\ \emph {et~al.}(2013)\citenamefont
  {Cameron}, \citenamefont {Wilson}, \citenamefont {Bernstein},\ and\
  \citenamefont {Kerfeld}}]{Cameron:2013wa}%
  \BibitemOpen
  \bibfield  {author} {\bibinfo {author} {\bibfnamefont {J.~C.}\ \bibnamefont
  {Cameron}}, \bibinfo {author} {\bibfnamefont {S.~C.}\ \bibnamefont {Wilson}},
  \bibinfo {author} {\bibfnamefont {S.~L.}\ \bibnamefont {Bernstein}}, \ and\
  \bibinfo {author} {\bibfnamefont {C.~A.}\ \bibnamefont {Kerfeld}},\ }\href
  {\doibase 10.1016/j.cell.2013.10.044} {\bibfield  {journal} {\bibinfo
  {journal} {Cell}\ }\textbf {\bibinfo {volume} {155}},\ \bibinfo {pages}
  {1131} (\bibinfo {year} {2013})}\BibitemShut {NoStop}%
\bibitem [{\citenamefont {Tanaka}\ \emph {et~al.}(2008)\citenamefont {Tanaka},
  \citenamefont {Kerfeld}, \citenamefont {Sawaya}, \citenamefont {Cai},
  \citenamefont {Heinhorst}, \citenamefont {Cannon},\ and\ \citenamefont
  {Yeates}}]{Tanaka:2008wp}%
  \BibitemOpen
  \bibfield  {author} {\bibinfo {author} {\bibfnamefont {S.}~\bibnamefont
  {Tanaka}}, \bibinfo {author} {\bibfnamefont {C.~A.}\ \bibnamefont {Kerfeld}},
  \bibinfo {author} {\bibfnamefont {M.~R.}\ \bibnamefont {Sawaya}}, \bibinfo
  {author} {\bibfnamefont {F.}~\bibnamefont {Cai}}, \bibinfo {author}
  {\bibfnamefont {S.}~\bibnamefont {Heinhorst}}, \bibinfo {author}
  {\bibfnamefont {G.~C.}\ \bibnamefont {Cannon}}, \ and\ \bibinfo {author}
  {\bibfnamefont {T.~O.}\ \bibnamefont {Yeates}},\ }\href {\doibase
  10.1126/science.1151458} {\bibfield  {journal} {\bibinfo  {journal}
  {Science}\ }\textbf {\bibinfo {volume} {319}},\ \bibinfo {pages} {1083}
  (\bibinfo {year} {2008})}\BibitemShut {NoStop}%
\bibitem [{\citenamefont {Lassila}\ \emph {et~al.}(2014)\citenamefont
  {Lassila}, \citenamefont {Bernstein}, \citenamefont {Kinney}, \citenamefont
  {Axen},\ and\ \citenamefont {Kerfeld}}]{Lassila:2014co}%
  \BibitemOpen
  \bibfield  {author} {\bibinfo {author} {\bibfnamefont {J.~K.}\ \bibnamefont
  {Lassila}}, \bibinfo {author} {\bibfnamefont {S.~L.}\ \bibnamefont
  {Bernstein}}, \bibinfo {author} {\bibfnamefont {J.~N.}\ \bibnamefont
  {Kinney}}, \bibinfo {author} {\bibfnamefont {S.~D.}\ \bibnamefont {Axen}}, \
  and\ \bibinfo {author} {\bibfnamefont {C.~A.}\ \bibnamefont {Kerfeld}},\
  }\href {\doibase 10.1016/j.jmb.2014.02.025} {\bibfield  {journal} {\bibinfo
  {journal} {J. Mol. Biol.}\ }\textbf {\bibinfo {volume} {426}},\ \bibinfo
  {pages} {2217} (\bibinfo {year} {2014})}\BibitemShut {NoStop}%
\bibitem [{\citenamefont {Savage}\ \emph {et~al.}(2010)\citenamefont {Savage},
  \citenamefont {Afonso}, \citenamefont {Chen},\ and\ \citenamefont
  {Silver}}]{Savage:2010ut}%
  \BibitemOpen
  \bibfield  {author} {\bibinfo {author} {\bibfnamefont {D.~F.}\ \bibnamefont
  {Savage}}, \bibinfo {author} {\bibfnamefont {B.}~\bibnamefont {Afonso}},
  \bibinfo {author} {\bibfnamefont {A.~H.}\ \bibnamefont {Chen}}, \ and\
  \bibinfo {author} {\bibfnamefont {P.~A.}\ \bibnamefont {Silver}},\ }\href
  {\doibase 10.1126/science.1186090} {\bibfield  {journal} {\bibinfo  {journal}
  {Science}\ }\textbf {\bibinfo {volume} {327}},\ \bibinfo {pages} {1258}
  (\bibinfo {year} {2010})}\BibitemShut {NoStop}%
\bibitem [{\citenamefont {Yeates}\ \emph {et~al.}(2007)\citenamefont {Yeates},
  \citenamefont {Tsai}, \citenamefont {Tanaka}, \citenamefont {Sawaya},\ and\
  \citenamefont {Kerfeld}}]{Yeates:2007cp}%
  \BibitemOpen
  \bibfield  {author} {\bibinfo {author} {\bibfnamefont {T.~O.}\ \bibnamefont
  {Yeates}}, \bibinfo {author} {\bibfnamefont {Y.}~\bibnamefont {Tsai}},
  \bibinfo {author} {\bibfnamefont {S.}~\bibnamefont {Tanaka}}, \bibinfo
  {author} {\bibfnamefont {M.~R.}\ \bibnamefont {Sawaya}}, \ and\ \bibinfo
  {author} {\bibfnamefont {C.~A.}\ \bibnamefont {Kerfeld}},\ }\href {\doibase
  10.1042/BST0350508} {\bibfield  {journal} {\bibinfo  {journal} {Biochem. Soc.
  Trans.}\ }\textbf {\bibinfo {volume} {35}},\ \bibinfo {pages} {508} (\bibinfo
  {year} {2007})}\BibitemShut {NoStop}%
\bibitem [{\citenamefont {Bobik}\ \emph {et~al.}(2015)\citenamefont {Bobik},
  \citenamefont {Lehman},\ and\ \citenamefont {Yeates}}]{Bobik:2015we}%
  \BibitemOpen
  \bibfield  {author} {\bibinfo {author} {\bibfnamefont {T.~A.}\ \bibnamefont
  {Bobik}}, \bibinfo {author} {\bibfnamefont {B.~P.}\ \bibnamefont {Lehman}}, \
  and\ \bibinfo {author} {\bibfnamefont {T.~O.}\ \bibnamefont {Yeates}},\
  }\href {\doibase 10.1111/mmi.13117} {\bibfield  {journal} {\bibinfo
  {journal} {Mol. Microbiol.}\ }\textbf {\bibinfo {volume} {98}},\ \bibinfo
  {pages} {193} (\bibinfo {year} {2015})}\BibitemShut {NoStop}%
\bibitem [{\citenamefont {Yeates}\ \emph {et~al.}(2010)\citenamefont {Yeates},
  \citenamefont {Crowley},\ and\ \citenamefont {Tanaka}}]{Yeates:2010kx}%
  \BibitemOpen
  \bibfield  {author} {\bibinfo {author} {\bibfnamefont {T.~O.}\ \bibnamefont
  {Yeates}}, \bibinfo {author} {\bibfnamefont {C.~S.}\ \bibnamefont {Crowley}},
  \ and\ \bibinfo {author} {\bibfnamefont {S.}~\bibnamefont {Tanaka}},\ }\href
  {\doibase 10.1146/annurev.biophys.093008.131418} {\bibfield  {journal}
  {\bibinfo  {journal} {Annu. Rev. Biophys.}\ }\textbf {\bibinfo {volume}
  {39}},\ \bibinfo {pages} {185} (\bibinfo {year} {2010})}\BibitemShut
  {NoStop}%
\bibitem [{\citenamefont {Cai}\ \emph {et~al.}(2015)\citenamefont {Cai},
  \citenamefont {Sutter}, \citenamefont {Bernstein}, \citenamefont {Kinney},\
  and\ \citenamefont {Kerfeld}}]{Cai:2015bf}%
  \BibitemOpen
  \bibfield  {author} {\bibinfo {author} {\bibfnamefont {F.}~\bibnamefont
  {Cai}}, \bibinfo {author} {\bibfnamefont {M.}~\bibnamefont {Sutter}},
  \bibinfo {author} {\bibfnamefont {S.~L.}\ \bibnamefont {Bernstein}}, \bibinfo
  {author} {\bibfnamefont {J.~N.}\ \bibnamefont {Kinney}}, \ and\ \bibinfo
  {author} {\bibfnamefont {C.~A.}\ \bibnamefont {Kerfeld}},\ }\href {\doibase
  10.1021/sb500226j} {\bibfield  {journal} {\bibinfo  {journal} {ACS Synth.
  Biol.}\ }\textbf {\bibinfo {volume} {4}},\ \bibinfo {pages} {444} (\bibinfo
  {year} {2015})}\BibitemShut {NoStop}%
\bibitem [{\citenamefont {Frey}\ \emph {et~al.}(2016)\citenamefont {Frey},
  \citenamefont {Mantri}, \citenamefont {Rocca},\ and\ \citenamefont
  {Hilvert}}]{Frey:2016cn}%
  \BibitemOpen
  \bibfield  {author} {\bibinfo {author} {\bibfnamefont {R.}~\bibnamefont
  {Frey}}, \bibinfo {author} {\bibfnamefont {S.}~\bibnamefont {Mantri}},
  \bibinfo {author} {\bibfnamefont {M.}~\bibnamefont {Rocca}}, \ and\ \bibinfo
  {author} {\bibfnamefont {D.}~\bibnamefont {Hilvert}},\ }\href {\doibase
  10.1021/jacs.6b04744} {\bibfield  {journal} {\bibinfo  {journal} {J. Am.
  Chem. Soc.}\ }\textbf {\bibinfo {volume} {138}},\ \bibinfo {pages} {10072}
  (\bibinfo {year} {2016})}\BibitemShut {NoStop}%
\bibitem [{\citenamefont {Giessen}\ and\ \citenamefont
  {Silver}(2017)}]{Giessen:2017ia}%
  \BibitemOpen
  \bibfield  {author} {\bibinfo {author} {\bibfnamefont {T.~W.}\ \bibnamefont
  {Giessen}}\ and\ \bibinfo {author} {\bibfnamefont {P.~A.}\ \bibnamefont
  {Silver}},\ }\href {\doibase 10.1016/j.copbio.2017.01.004} {\bibfield
  {journal} {\bibinfo  {journal} {Curr. Opin. Biotechnol.}\ }\textbf {\bibinfo
  {volume} {46}},\ \bibinfo {pages} {42} (\bibinfo {year} {2017})}\BibitemShut
  {NoStop}%
\bibitem [{\citenamefont {Chen}\ \emph {et~al.}(2006)\citenamefont {Chen},
  \citenamefont {Daniel}, \citenamefont {Quinkert}, \citenamefont {De},
  \citenamefont {Stein}, \citenamefont {Bowman}, \citenamefont {Chipman},
  \citenamefont {Rotello}, \citenamefont {Kao},\ and\ \citenamefont
  {Dragnea}}]{Chen:2006kj}%
  \BibitemOpen
  \bibfield  {author} {\bibinfo {author} {\bibfnamefont {C.}~\bibnamefont
  {Chen}}, \bibinfo {author} {\bibfnamefont {M.-C.}\ \bibnamefont {Daniel}},
  \bibinfo {author} {\bibfnamefont {Z.~T.}\ \bibnamefont {Quinkert}}, \bibinfo
  {author} {\bibfnamefont {M.}~\bibnamefont {De}}, \bibinfo {author}
  {\bibfnamefont {B.}~\bibnamefont {Stein}}, \bibinfo {author} {\bibfnamefont
  {V.~D.}\ \bibnamefont {Bowman}}, \bibinfo {author} {\bibfnamefont {P.~R.}\
  \bibnamefont {Chipman}}, \bibinfo {author} {\bibfnamefont {V.~M.}\
  \bibnamefont {Rotello}}, \bibinfo {author} {\bibfnamefont {C.~C.}\
  \bibnamefont {Kao}}, \ and\ \bibinfo {author} {\bibfnamefont
  {B.}~\bibnamefont {Dragnea}},\ }\href {\doibase 10.1021/nl0600878} {\bibfield
   {journal} {\bibinfo  {journal} {Nano Lett.}\ }\textbf {\bibinfo {volume}
  {6}},\ \bibinfo {pages} {611} (\bibinfo {year} {2006})}\BibitemShut {NoStop}%
\bibitem [{\citenamefont {Sutter}\ \emph {et~al.}(2017)\citenamefont {Sutter},
  \citenamefont {Greber}, \citenamefont {Aussignargues},\ and\ \citenamefont
  {Kerfeld}}]{Sutter:2017ew}%
  \BibitemOpen
  \bibfield  {author} {\bibinfo {author} {\bibfnamefont {M.}~\bibnamefont
  {Sutter}}, \bibinfo {author} {\bibfnamefont {B.}~\bibnamefont {Greber}},
  \bibinfo {author} {\bibfnamefont {C.}~\bibnamefont {Aussignargues}}, \ and\
  \bibinfo {author} {\bibfnamefont {C.~A.}\ \bibnamefont {Kerfeld}},\ }\href
  {\doibase 10.1126/science.aan3289} {\bibfield  {journal} {\bibinfo  {journal}
  {Science}\ }\textbf {\bibinfo {volume} {356}},\ \bibinfo {pages} {1293}
  (\bibinfo {year} {2017})}\BibitemShut {NoStop}%
\bibitem [{\citenamefont {Cai}\ \emph {et~al.}(2009)\citenamefont {Cai},
  \citenamefont {Menon}, \citenamefont {Cannon}, \citenamefont {Curry},
  \citenamefont {Shively},\ and\ \citenamefont {Heinhorst}}]{Cai:2009et}%
  \BibitemOpen
  \bibfield  {author} {\bibinfo {author} {\bibfnamefont {F.}~\bibnamefont
  {Cai}}, \bibinfo {author} {\bibfnamefont {B.~B.}\ \bibnamefont {Menon}},
  \bibinfo {author} {\bibfnamefont {G.~C.}\ \bibnamefont {Cannon}}, \bibinfo
  {author} {\bibfnamefont {K.~J.}\ \bibnamefont {Curry}}, \bibinfo {author}
  {\bibfnamefont {J.~M.}\ \bibnamefont {Shively}}, \ and\ \bibinfo {author}
  {\bibfnamefont {S.}~\bibnamefont {Heinhorst}},\ }\href {\doibase
  10.1371/journal.pone.0007521} {\bibfield  {journal} {\bibinfo  {journal}
  {PLOS ONE}\ }\textbf {\bibinfo {volume} {4}},\ \bibinfo {pages} {e7521}
  (\bibinfo {year} {2009})}\BibitemShut {NoStop}%
\bibitem [{\citenamefont {Iancu}\ \emph {et~al.}(2010)\citenamefont {Iancu},
  \citenamefont {Morris}, \citenamefont {Dou}, \citenamefont {Heinhorst},
  \citenamefont {Cannon},\ and\ \citenamefont {Jensen}}]{Iancu:2010eb}%
  \BibitemOpen
  \bibfield  {author} {\bibinfo {author} {\bibfnamefont {C.~V.}\ \bibnamefont
  {Iancu}}, \bibinfo {author} {\bibfnamefont {D.~M.}\ \bibnamefont {Morris}},
  \bibinfo {author} {\bibfnamefont {Z.}~\bibnamefont {Dou}}, \bibinfo {author}
  {\bibfnamefont {S.}~\bibnamefont {Heinhorst}}, \bibinfo {author}
  {\bibfnamefont {G.~C.}\ \bibnamefont {Cannon}}, \ and\ \bibinfo {author}
  {\bibfnamefont {G.~J.}\ \bibnamefont {Jensen}},\ }\href {\doibase
  10.1016/j.jmb.2009.11.019} {\bibfield  {journal} {\bibinfo  {journal} {J.
  Mol. Biol.}\ }\textbf {\bibinfo {volume} {396}},\ \bibinfo {pages} {105}
  (\bibinfo {year} {2010})}\BibitemShut {NoStop}%
\bibitem [{\citenamefont {Perlmutter}\ \emph {et~al.}(2016)\citenamefont
  {Perlmutter}, \citenamefont {Mohajerani},\ and\ \citenamefont
  {Hagan}}]{Perlmutter:2016hz}%
  \BibitemOpen
  \bibfield  {author} {\bibinfo {author} {\bibfnamefont {J.~D.}\ \bibnamefont
  {Perlmutter}}, \bibinfo {author} {\bibfnamefont {F.}~\bibnamefont
  {Mohajerani}}, \ and\ \bibinfo {author} {\bibfnamefont {M.~F.}\ \bibnamefont
  {Hagan}},\ }\href {\doibase 10.7554/eLife.14078} {\bibfield  {journal}
  {\bibinfo  {journal} {eLife}\ ,\ \bibinfo {pages} {5:e14078}} (\bibinfo
  {year} {2016})}\BibitemShut {NoStop}%
\bibitem [{\citenamefont {Wagner}\ and\ \citenamefont
  {Zandi}(2015)}]{Wagner:2015jy}%
  \BibitemOpen
  \bibfield  {author} {\bibinfo {author} {\bibfnamefont {J.}~\bibnamefont
  {Wagner}}\ and\ \bibinfo {author} {\bibfnamefont {R.}~\bibnamefont {Zandi}},\
  }\href {\doibase 10.1016/j.bpj.2015.07.041} {\bibfield  {journal} {\bibinfo
  {journal} {Biophys. J}\ }\textbf {\bibinfo {volume} {109}},\ \bibinfo {pages}
  {956} (\bibinfo {year} {2015})}\BibitemShut {NoStop}%
\bibitem [{\citenamefont {Hicks}\ and\ \citenamefont
  {Henley}(2006)}]{Hicks:2006gc}%
  \BibitemOpen
  \bibfield  {author} {\bibinfo {author} {\bibfnamefont {S.~D.}\ \bibnamefont
  {Hicks}}\ and\ \bibinfo {author} {\bibfnamefont {C.~L.}\ \bibnamefont
  {Henley}},\ }\href {\doibase 10.1103/PhysRevE.74.031912} {\bibfield
  {journal} {\bibinfo  {journal} {Phys. Rev. E}\ }\textbf {\bibinfo {volume}
  {74}},\ \bibinfo {pages} {031912} (\bibinfo {year} {2006})}\BibitemShut
  {NoStop}%
\bibitem [{\citenamefont {Sutter}\ \emph {et~al.}(2016)\citenamefont {Sutter},
  \citenamefont {Faulkner}, \citenamefont {Aussignargues}, \citenamefont
  {Paasch}, \citenamefont {Barrett}, \citenamefont {Kerfeld},\ and\
  \citenamefont {Liu}}]{Sutter:2016}%
  \BibitemOpen
  \bibfield  {author} {\bibinfo {author} {\bibfnamefont {M.}~\bibnamefont
  {Sutter}}, \bibinfo {author} {\bibfnamefont {M.}~\bibnamefont {Faulkner}},
  \bibinfo {author} {\bibfnamefont {C.}~\bibnamefont {Aussignargues}}, \bibinfo
  {author} {\bibfnamefont {B.~C.}\ \bibnamefont {Paasch}}, \bibinfo {author}
  {\bibfnamefont {S.}~\bibnamefont {Barrett}}, \bibinfo {author} {\bibfnamefont
  {C.~A.}\ \bibnamefont {Kerfeld}}, \ and\ \bibinfo {author} {\bibfnamefont
  {L.-N.}\ \bibnamefont {Liu}},\ }\href {\doibase 10.1021/acs.nanolett.5b04259}
  {\bibfield  {journal} {\bibinfo  {journal} {Nano Letters}\ }\textbf {\bibinfo
  {volume} {16}},\ \bibinfo {pages} {1590} (\bibinfo {year}
  {2016})}\BibitemShut {NoStop}%
\bibitem [{\citenamefont {Seung}\ and\ \citenamefont
  {Nelson}(1988)}]{Seung:1988ix}%
  \BibitemOpen
  \bibfield  {author} {\bibinfo {author} {\bibfnamefont {H.~S.}\ \bibnamefont
  {Seung}}\ and\ \bibinfo {author} {\bibfnamefont {D.~R.}\ \bibnamefont
  {Nelson}},\ }\href {\doibase 10.1103/PhysRevA.38.1005} {\bibfield  {journal}
  {\bibinfo  {journal} {Phys. Rev. A}\ }\textbf {\bibinfo {volume} {38}},\
  \bibinfo {pages} {1005} (\bibinfo {year} {1988})}\BibitemShut {NoStop}%
\bibitem [{\citenamefont {Kinney}\ \emph {et~al.}(2011)\citenamefont {Kinney},
  \citenamefont {Axen},\ and\ \citenamefont {Kerfeld}}]{Kinney:2011ps}%
  \BibitemOpen
  \bibfield  {author} {\bibinfo {author} {\bibfnamefont {J.~N.}\ \bibnamefont
  {Kinney}}, \bibinfo {author} {\bibfnamefont {S.~D.}\ \bibnamefont {Axen}}, \
  and\ \bibinfo {author} {\bibfnamefont {C.~A.}\ \bibnamefont {Kerfeld}},\
  }\href {\doibase 10.1007/s11120-011-9624-6} {\bibfield  {journal} {\bibinfo
  {journal} {Photosynth. Res.}\ }\textbf {\bibinfo {volume} {109}},\ \bibinfo
  {pages} {21} (\bibinfo {year} {2011})}\BibitemShut {NoStop}%
\bibitem [{\citenamefont {Bahrami}\ \emph {et~al.}(2012)\citenamefont
  {Bahrami}, \citenamefont {Lipowsky},\ and\ \citenamefont
  {Weikl}}]{Bahrami:2012gb}%
  \BibitemOpen
  \bibfield  {author} {\bibinfo {author} {\bibfnamefont {A.~H.}\ \bibnamefont
  {Bahrami}}, \bibinfo {author} {\bibfnamefont {R.}~\bibnamefont {Lipowsky}}, \
  and\ \bibinfo {author} {\bibfnamefont {T.~R.}\ \bibnamefont {Weikl}},\ }\href
  {\doibase 10.1103/PhysRevLett.109.188102} {\bibfield  {journal} {\bibinfo
  {journal} {Phys. Rev. Lett.}\ }\textbf {\bibinfo {volume} {109}},\ \bibinfo
  {pages} {188102} (\bibinfo {year} {2012})}\BibitemShut {NoStop}%
\bibitem [{\citenamefont {van~der Wel}\ \emph {et~al.}(2016)\citenamefont
  {van~der Wel}, \citenamefont {Vahid}, \citenamefont {{\v{S}}ari{\'c}},
  \citenamefont {Idema}, \citenamefont {Heinrich},\ and\ \citenamefont
  {Kraft}}]{vanderWel:2016du}%
  \BibitemOpen
  \bibfield  {author} {\bibinfo {author} {\bibfnamefont {C.}~\bibnamefont
  {van~der Wel}}, \bibinfo {author} {\bibfnamefont {A.}~\bibnamefont {Vahid}},
  \bibinfo {author} {\bibfnamefont {A.}~\bibnamefont {{\v{S}}ari{\'c}}},
  \bibinfo {author} {\bibfnamefont {T.}~\bibnamefont {Idema}}, \bibinfo
  {author} {\bibfnamefont {D.}~\bibnamefont {Heinrich}}, \ and\ \bibinfo
  {author} {\bibfnamefont {D.~J.}\ \bibnamefont {Kraft}},\ }\href {\doibase
  10.1038/srep32825} {\bibfield  {journal} {\bibinfo  {journal} {Sci. Rep.}\
  }\textbf {\bibinfo {volume} {6}},\ \bibinfo {pages} {srep32825} (\bibinfo
  {year} {2016})}\BibitemShut {NoStop}%
\bibitem [{\citenamefont {{\v{S}}ari{\'c}}\ and\ \citenamefont
  {Cacciuto}(2012)}]{Saric:2012hb}%
  \BibitemOpen
  \bibfield  {author} {\bibinfo {author} {\bibfnamefont {A.}~\bibnamefont
  {{\v{S}}ari{\'c}}}\ and\ \bibinfo {author} {\bibfnamefont {A.}~\bibnamefont
  {Cacciuto}},\ }\href {\doibase 10.1103/PhysRevLett.109.188101} {\bibfield
  {journal} {\bibinfo  {journal} {Phys. Rev. Lett.}\ }\textbf {\bibinfo
  {volume} {109}},\ \bibinfo {pages} {188101} (\bibinfo {year}
  {2012})}\BibitemShut {NoStop}%
\bibitem [{\citenamefont {Funkhouser}\ \emph {et~al.}(2013)\citenamefont
  {Funkhouser}, \citenamefont {Sknepnek},\ and\ \citenamefont {Olvera de~la
  Cruz}}]{Funkhouser:2012hea}%
  \BibitemOpen
  \bibfield  {author} {\bibinfo {author} {\bibfnamefont {C.~M.}\ \bibnamefont
  {Funkhouser}}, \bibinfo {author} {\bibfnamefont {R.}~\bibnamefont
  {Sknepnek}}, \ and\ \bibinfo {author} {\bibfnamefont {M.}~\bibnamefont
  {Olvera de~la Cruz}},\ }\href {\doibase 10.1039/C2SM26607E} {\bibfield
  {journal} {\bibinfo  {journal} {Soft Matter}\ }\textbf {\bibinfo {volume}
  {9}},\ \bibinfo {pages} {60} (\bibinfo {year} {2013})}\BibitemShut {NoStop}%
\bibitem [{\citenamefont {Bowick}\ \emph {et~al.}(2002)\citenamefont {Bowick},
  \citenamefont {Cacciuto}, \citenamefont {Nelson},\ and\ \citenamefont
  {Travesset}}]{Bowick:2002cf}%
  \BibitemOpen
  \bibfield  {author} {\bibinfo {author} {\bibfnamefont {M.}~\bibnamefont
  {Bowick}}, \bibinfo {author} {\bibfnamefont {A.}~\bibnamefont {Cacciuto}},
  \bibinfo {author} {\bibfnamefont {D.~R.}\ \bibnamefont {Nelson}}, \ and\
  \bibinfo {author} {\bibfnamefont {A.}~\bibnamefont {Travesset}},\ }\href
  {\doibase 10.1103/PhysRevLett.89.185502} {\bibfield  {journal} {\bibinfo
  {journal} {Phys. Rev. Lett.}\ }\textbf {\bibinfo {volume} {89}},\ \bibinfo
  {pages} {185502} (\bibinfo {year} {2002})}\BibitemShut {NoStop}%
\bibitem [{\citenamefont {Stratford}\ \emph {et~al.}(2005)\citenamefont
  {Stratford}, \citenamefont {Adhikari}, \citenamefont {Pagonabarraga},
  \citenamefont {Desplat},\ and\ \citenamefont {Cates}}]{Stratford:2005}%
  \BibitemOpen
  \bibfield  {author} {\bibinfo {author} {\bibfnamefont {K.}~\bibnamefont
  {Stratford}}, \bibinfo {author} {\bibfnamefont {R.}~\bibnamefont {Adhikari}},
  \bibinfo {author} {\bibfnamefont {I.}~\bibnamefont {Pagonabarraga}}, \bibinfo
  {author} {\bibfnamefont {J.-C.}\ \bibnamefont {Desplat}}, \ and\ \bibinfo
  {author} {\bibfnamefont {M.~E.}\ \bibnamefont {Cates}},\ }\href {\doibase
  10.1126/science.1116589} {\bibfield  {journal} {\bibinfo  {journal}
  {Science}\ }\textbf {\bibinfo {volume} {309}},\ \bibinfo {pages} {2198}
  (\bibinfo {year} {2005})}\BibitemShut {NoStop}%
\bibitem [{\citenamefont {Chandler}(1987)}]{Chandler:1987tp}%
  \BibitemOpen
  \bibfield  {author} {\bibinfo {author} {\bibfnamefont {D.}~\bibnamefont
  {Chandler}},\ }\href@noop {} {{\emph {\bibinfo
  {title} {{Introduction to Modern Statistical Mechanics}}}}}\ (\bibinfo
  {publisher} {Oxford University Press, USA},\ \bibinfo {year}
  {1987})\BibitemShut {NoStop}%
\bibitem [{\citenamefont {Humphrey}\ \emph {et~al.}(1996)\citenamefont
  {Humphrey}, \citenamefont {Dalke},\ and\ \citenamefont {Schulten}}]{vmd}%
  \BibitemOpen
  \bibfield  {author} {\bibinfo {author} {\bibfnamefont {W.}~\bibnamefont
  {Humphrey}}, \bibinfo {author} {\bibfnamefont {A.}~\bibnamefont {Dalke}}, \
  and\ \bibinfo {author} {\bibfnamefont {K.}~\bibnamefont {Schulten}},\
  }\href@noop {} {\bibfield  {journal} {\bibinfo  {journal} {J. Mol. Graphics}\
  }\textbf {\bibinfo {volume} {14}},\ \bibinfo {pages} {33} (\bibinfo {year}
  {1996})}\BibitemShut {NoStop}%
\bibitem [{\citenamefont {Steinhardt}\ \emph {et~al.}(1983)\citenamefont
  {Steinhardt}, \citenamefont {Nelson},\ and\ \citenamefont
  {Ronchetti}}]{Steinhardt:1983uh}%
  \BibitemOpen
  \bibfield  {author} {\bibinfo {author} {\bibfnamefont {P.}~\bibnamefont
  {Steinhardt}}, \bibinfo {author} {\bibfnamefont {D.}~\bibnamefont {Nelson}},
  \ and\ \bibinfo {author} {\bibfnamefont {M.}~\bibnamefont {Ronchetti}},\
  }\href {\doibase 10.1103/PhysRevB.28.784} {\bibfield  {journal} {\bibinfo
  {journal} {Phys. Rev. B}\ }\textbf {\bibinfo {volume} {28}},\ \bibinfo
  {pages} {784} (\bibinfo {year} {1983})}\BibitemShut {NoStop}%
\end{thebibliography}%

\clearpage

\onecolumngrid

\section*{Supplementary Materials}

\newcommand{\beginsupplement}{%
        \setcounter{table}{0}
        \renewcommand{\thetable}{S\arabic{table}}%
        \setcounter{figure}{0}
        \renewcommand{\thefigure}{S\arabic{figure}}%
        \setcounter{equation}{0}
        \renewcommand{\theequation}{S\arabic{equation}}%
        \setcounter{section}{0}
        \renewcommand{\thesection}{S\arabic{section}}%

     }

\beginsupplement

\begin{figure}[ht!!]
   \begin{center}
     \includegraphics[width=0.5\linewidth]{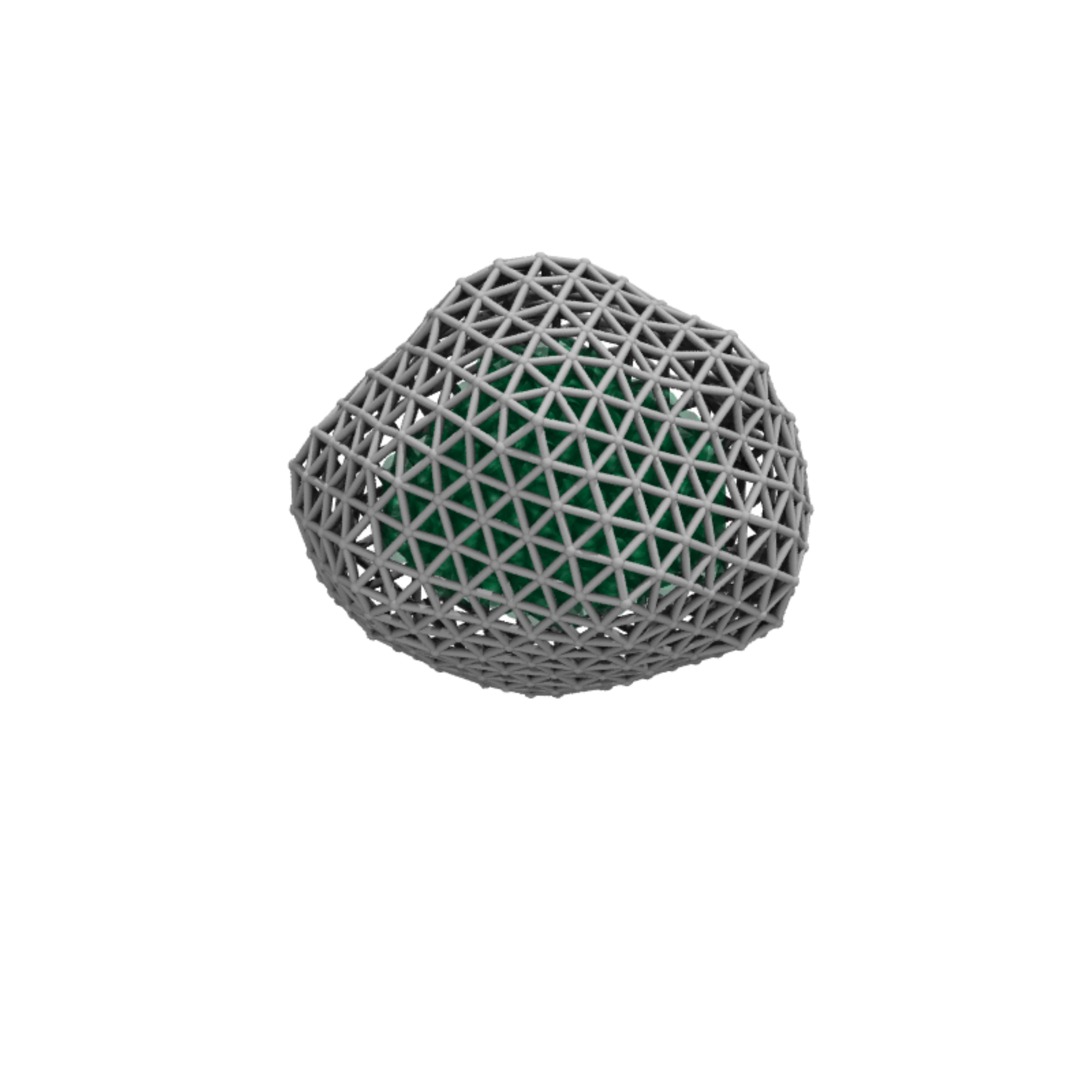}
   \end{center}
   \caption{ A still from the animation
     of an assembly trajectory is shown with the shell components in gray and the cargo as spheres centered at the cargo lattice sites. In order to clearly show the shell structure, only occupied lattice sites that are above a distance cutoff of 1.5$\l_0$ are shown in our visualizations. 
     }
\end{figure}

\section{Monte Carlo dynamics of the molecular model}\label{sec:mcmc}

The Monte Carlo dynamics of our molecular model for
microcompartment assembly involves several kinds of trial moves, each
constructed to be reversible and to preserve the grand canonical
probability distribution $p(X,N_{\textrm{s}}, \{\sigma\})$ in Eq. [5]. Here we detail these
moves and their associated acceptance probabilities. For the cargo we
require only a {\em cargo addition/removal} move.  For the shell we
employ: {\em vertex displacement} moves, {\em simple monomer
  insertion/deletion} moves, {\em wedge insertion/deletion} moves, and
{\em vertex fusion/fission} moves.

\subsection{Vertex distribution function}

Representing the protein shell as a triangulated sheet introduces a
subtlety to the grand canonical distribution function $p(X,N)$.
Triangles in the sheet each correspond to a molecular monomer with
$3\times 3$ degrees of freedom, corresponding to the positions of each
of its constituent vertices. Physically, a pair of such monomers would
possess $2\times 3\times 3$ degrees of freedom even when
bound. Represented as a ``sheet'' comprising two juxtaposed triangles
(with a total of four vertices), however, a bound dimer possesses only
$4\times 3$ degrees of freedom. The implied constraint mimics strong
and short-ranged forces of cohesion between bound monomers, which
limit the fluctuations of a bound vertex to a very small volume $v_a$
about its binding partner. For a favorable binding energy
$\epsilon_{\rm bind}$ that is constant within this volume, the binding
equilibrium for a pair of monomers is parameterized by the resulting
vertex affinity $K=v_a e^{-\beta\epsilon_{\rm bind}}$ ~\cite{Chandler:1987tp}.
In setting $v_a=0$, we implicitly take
$\epsilon_{\rm bind} \rightarrow -\infty$ such that $K$ remains
nonzero and finite.

The triangulated sheet representation does not resolve fluctuations
of bound vertices on the irrelevantly small length scale
$v_a^{1/3}$. These fluctuations have effectively been integrated out,
with each bound pair contributing a factor of $K$ to the probability
distribution for sheet vertices:
\begin{equation}
  \bar{p}(\bar{X},N_{\rm vertex},\{\sigma\}) = K^{N_{\rm bind}}z_{\textrm{s}}^{N_{\textrm{s}}}e^{-\beta({\cal H}_{\rm el} +
    {\cal H}_{\rm cargo} +{\cal H}_{\rm int} )}/\Xi,
\end{equation}
where $N_{\rm bind} = 3 N_{\textrm{s}} - N_{\rm vertex}$, and $\bar{X}$ denotes the configuration of the
set of $N_{\rm vertex}$ distinct vertices comprising the triangulated
sheet. The activity of shell monomers,
\begin{equation}
  z_{\textrm{s}} = e^{\beta \mu_{\textrm{s}}}/\lambda^9
  \label{eqn:zs}
\end{equation}
has units of (length)$^{-9}$ reflecting the 9 degrees of freedom for
an unbound monomer. As written in Eq.~\ref{eqn:zs}, $\lambda$ is an
arbitrary length scale that sets the compensatingly arbitrary zero of
chemical potential $\mu_{\textrm{s}}$.

\subsection{Metropolis acceptance probability}
Our simulations are designed to sample the probability distribution
$\bar{p}(\bar{X},N_{\rm vertex},\{\sigma\})$. In a Metropolis Monte
Carlo scheme, detailed balance with respect to this distribution is
ensured by accepting proposed moves with probability
\begin{equation}
  \textrm{acc}(\Gamma \to \Gamma') = 
  \min \left[ 1, \frac{\textrm{gen}(\Gamma' \to \Gamma) \bar{p}(\Gamma')}
    {\textrm{gen}(\Gamma \to \Gamma') \bar{p}(\Gamma')}
    \right],
\end{equation}
where $\Gamma$ is shorthand for the variables $(\bar{X},N_{\rm
  vertex},\{\sigma\})$ that define the system's microscopic state,
and $\textrm{gen}(\Gamma \to \Gamma')$ is the probability of proposing
$\Gamma'$ as a trial state from the existing state $\Gamma$.

Details of our Monte Carlo moves, and the resulting changes in
equilibrium probability $\bar{p}(\Gamma)$ that result, are described
below. For those that change the number of shell monomers, acceptance
probabilties involve factors of $z_{\textrm{s}}$ and $K$ in addition to the
energetics described by
${\cal H}_{\rm shell}$ and
${\cal H}_{\rm int}$.  Several of these trial moves have asymmetric
generation probabilities, $\textrm{gen}(\Gamma' \to \Gamma) \neq
\textrm{gen}(\Gamma \to \Gamma')$, which must be accounted for as well.

\subsection{Vertex displacement}

\begin{figure}
  \begin{center}
  \qquad
\includegraphics[width=0.5\linewidth]{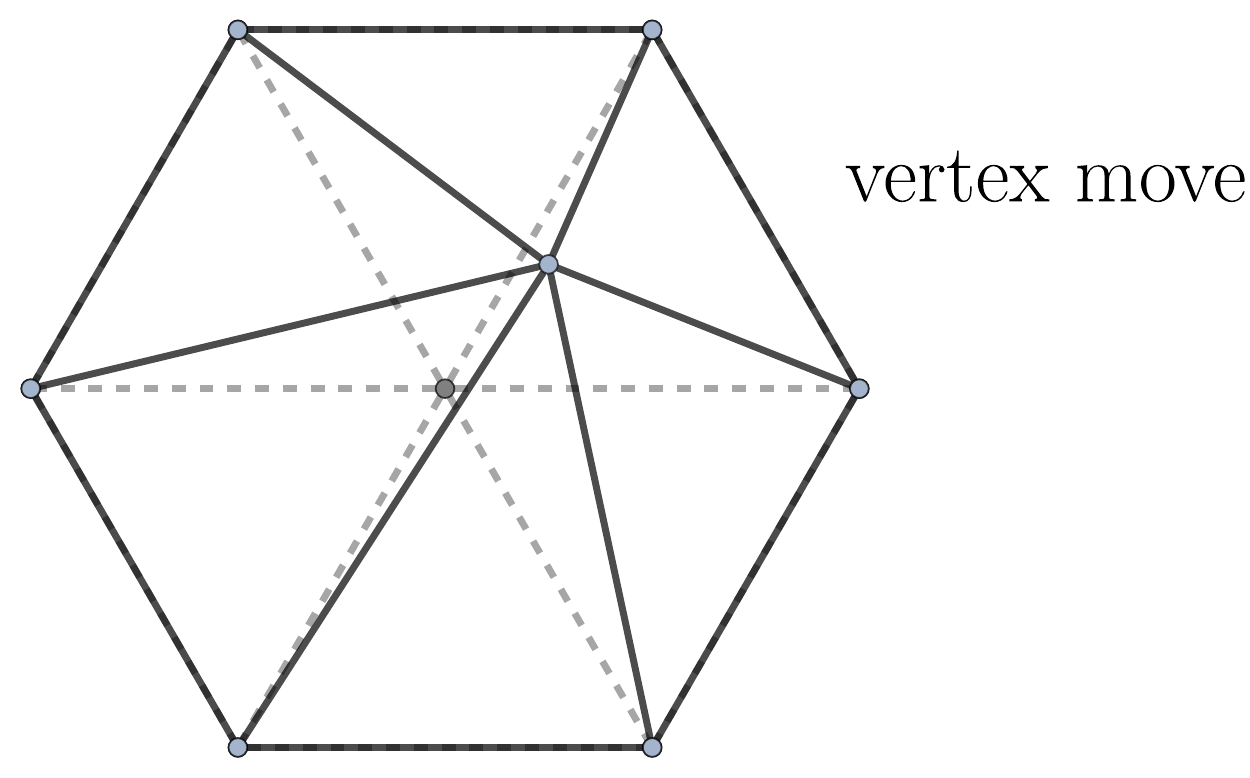}
\end{center}
  \caption{A vertex is randomly displaced and the energy
    difference between the initial and final states is computed. The
    move is then accepted or rejected so as to guarantee detailed
    balance as detailed in the text. }
\label{fig:vertexmove}
\end{figure}

Elastic relaxation of the shell is accomplished by conventional Monte
Carlo displacement moves. Such a move selects a vertex at random and
attempts to displace it by an amount $\Delta a$ that is chosen at random from a symmetric distribution, as depicted in Fig.~S2.
The standard acceptance criterion is given in Eq.~[4].

For a state $\Gamma$ with $N_{\rm vertex}(\Gamma)$ shell monomers, a
Monte Carlo sweep is defined as a sequence of $N_{\rm vertex}(\Gamma)$
displacement moves, each acting on a randomly chosen vertex.

\subsection{Simple monomer insertion/deletion}

\begin{table}
  \caption{
    Notation used to define the state of our molecular model and
    to formulate Monte Carlo moves that satisfy detailed balance.
  }
\begin{center}
\begin{tabular}{|l|c|}
\hline
  $\Gamma$ & a configuration \\ \hline
  $N_\textrm{vertex}$ & number of vertices of the shell structure \\ \hline
    $N_{\textrm{perim}}$ & number of perimeter edges of the shell \\ \hline
  $N_\textrm{cargo}$ & number of occupied cargo sites \\ \hline
  $N_{\textrm{boundary}}$ & number of boundary cargo site \\ \hline
    $v_{\textrm{add}}$ & spherical volume for monomer addition \\ \hline
      $N_{\textrm{fuse}}$ & number of possible fusion moves \\ \hline
        $N_{\textrm{fission}}$ & number of possible fission moves \\ \hline
            $v_{\textrm{fuse}}$ & spherical volume for fusion moves\\ \hline
\end{tabular}
\end{center}
\end{table}

\begin{figure}
  \begin{center}
\includegraphics[width=0.45\linewidth]{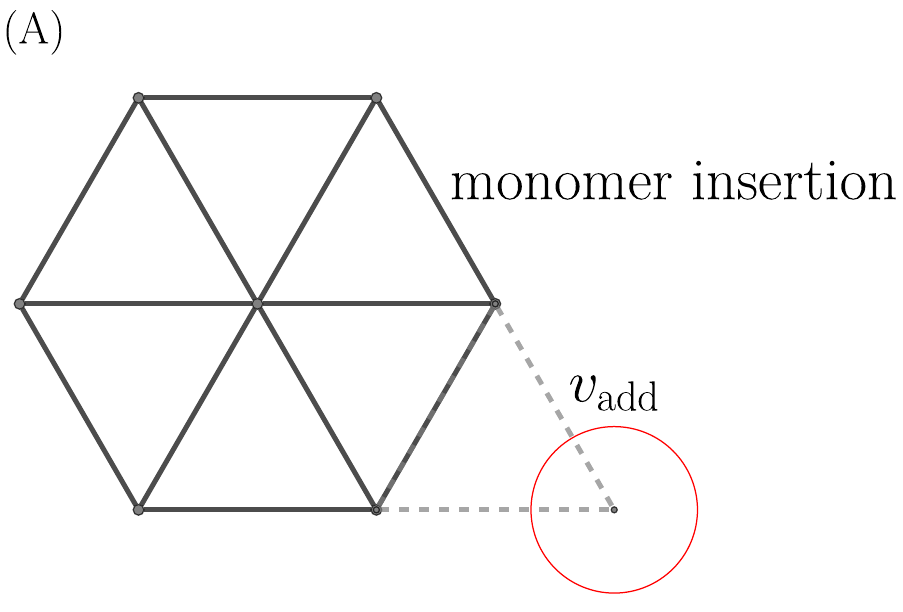}
\includegraphics[width=0.45\linewidth]{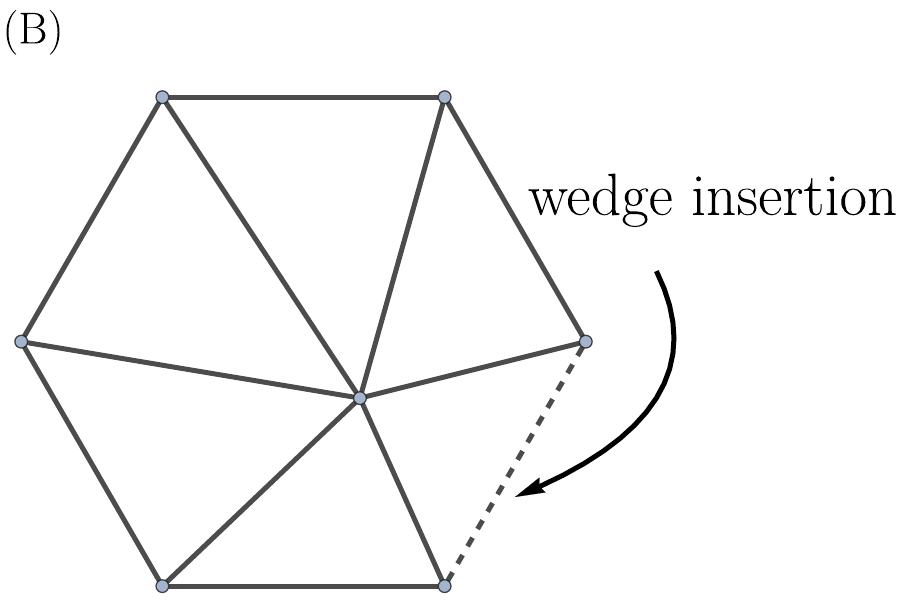}
\end{center}
  \caption{A schematic depiction of shell monomer insertion moves,
including simple monomer addition (A) and wedge insertion (B).
  }
	\label{fig:growthmoves}
\end{figure}

Our shell growth moves are depicted in Fig.~\ref{fig:growthmoves}.
They are typically attempted less than once per sweep, at a rate
proportional to the number $N_{\rm perim}$ of edges at the perimeter of
the sheet. Precisely, in each sweep we propose a shell monomer
insertion move with probability $N_{\rm perim} k_{\textrm{s}}^0 \tau$ (where $k_{\textrm{s}}^0 \tau$
is presumed to be much less than $1/N_{\rm perim}$).
First, one of the $N_{\rm perim}$ edges is
selected at random. Each of these edges corresponds to a particular
monomer $i$.  The type of growth move that is proposed depends on the
local connectivity and geometry of the sheet.

In a simple monomer insertion move (Fig.~\ref{fig:growthmoves} (A)), addition of a new shell monomer is
proposed by adding a single vertex that is connected to the two
existing vertices defining the selected edge. Because the vertices of
bound monomers are constrained to coincide exactly along the shared
edge, a new monomer defined by three vertices can be introduced by
adding just one new vertex.
The position of the new vertex is
generated by first identifying an ``ideal'' insertion site $a_{\rm
  ideal}$, lying in the same plane as monomer $i$, at a distance
$\sqrt{3}/3 l_0$ from the midpoint of the selected edge, and equidistant
from the edge's two endpoints.
The position $a$ of the new vertex is chosen randomly within a small
volume $v_\textrm{add}$ centered on $a_{\rm ideal}$.  The new monomer's outward
normal vector is chosen to have a positive projection onto that of
monomer $i$.  The generation probability for such a simple monomer
insertion is
\begin{equation}
  \textrm{gen}(\Gamma\to \Gamma') = N_{\rm perim}(\Gamma) k_{\textrm{s}}^0 \tau
  \frac{1}{v_\textrm{add}
    N_\textrm{perim}(\Gamma)}.
\end{equation}

The reverse of a simple addition move deletes a shell monomer by
removing a single vertex and its two edges. We first select one of the
$N_\textrm{perim}(\Gamma')$ edges at the perimeter of the sheet.
 A monomer added by simple insertion can be removed by selecting either
of its two exposed edges.
The generation probability for simple deletion is therefore
\begin{equation}
  \textrm{gen}(\Gamma' \to \Gamma) = N_{\rm perim}(\Gamma') k_{\textrm{s}}^0 \tau
  \frac{2}{
    N_\textrm{perim}(\Gamma')}.
\end{equation}

The grand canonical probability of $\Gamma'$ relative to
the initial configuration is
\begin{equation}
  \frac{\bar{p}(\Gamma')}{\bar{p}(\Gamma)} = z_{\textrm{s}} K^2
  \exp \left( -\beta \Delta {\cal H} \right),
\end{equation}
The two factors of $K$ account for constraining positions of two of
the vertices defining the new monomer.  The energy difference $\Delta
{\cal H}$ due to changes in
${\cal H}_{\rm shell}+{\cal H}_{\rm int}$
involves
only the elastic energy of the edges of the new monomer and the
interaction with nearby cargo lattice sites.

The Metropolis acceptance criterion for the simple monomer insertion
move is thus
\begin{equation}
  \textrm{acc}(\Gamma\to \Gamma') = \min\left[1, 2 z_{\textrm{s}} K^2
    v_\textrm{add}
    e^{-\beta \Delta {\cal H}} \right].
\end{equation}
In the acceptance probability for simple monomer deletion,
$\textrm{acc}(\Gamma'\to \Gamma)$, the second argument of the min
function is inverted.

Some surface edges are not eligible for simple insertion in our
protocol. Specifically, in many configurations it is possible to add a
shell monomer by filling in a ``wedge'' at the sheet's perimeter, as
depicted in Fig.~\ref{fig:growthmoves} (B).  We do not attempt simple insertion moves at the
edges defining such a wedge. When such an edge is selected, we instead
propose the wedge insertion move described below.

\subsection{Wedge insertion/deletion}

In some shell configurations a new monomer can be added without
introducing any new vertices at all. Geometrically, this corresponds
to filling in an empty wedge at the sheet's boundary, as sketched in
Fig~\ref{fig:growthmoves} (B).
We detect this possibility by checking
if two disconnected perimeter vertices share a neighbor and are within a distance $l<l_\textrm{max}$. Since the new
monomer is completely constrained in this case, the corresponding
generation probability is
\begin{equation}
  \textrm{gen}(\Gamma\to \Gamma') = N_{\rm perim}(\Gamma) k_{\textrm{s}}^0 \tau
  \frac{2}{N_\textrm{perim}(\Gamma)},
\end{equation}
whose factor of 2 reflects that the same insertion can result from
selecting either of the two existing edges that define the wedge.
The generation probability for the reverse move is simply
\begin{equation}
  \textrm{gen}(\Gamma' \to \Gamma) = N_{\rm perim}(\Gamma') k_{\textrm{s}}^0 \tau
  \frac{1}{N_\textrm{perim}(\Gamma')},
\end{equation}

The grand canonical probability of $\Gamma'$ relative to
the initial configuration is
\begin{equation}
  \frac{\bar{p}(\Gamma')}{\bar{p}(\Gamma)} = z_{\textrm{s}} K^3
  \exp \left( -\beta \Delta {\cal H} \right).
\end{equation}
As with the simple insertion move, factors of $K$ account for
constraining positions of vertices defining the new monomer.  The
energy difference $\Delta {\cal H}$ due to changes in ${\cal H}_{\rm
  shell}+{\cal H}_{\rm int}$ again involves only volume exclusion, the
elastic energy of the edges of the new monomer, and the interaction
with nearby cargo lattice sites.

The resulting Metropolis acceptance criterion for wedge insertion
is
\begin{equation}
  \textrm{acc}(\Gamma\to \Gamma') = \min\left[1,\frac{z_{\textrm{s}} K^3}{2} 
    e^{-\beta \Delta {\cal H}} \right].
\end{equation}
In the acceptance probability for wedge monomer deletion,
$\textrm{acc}(\Gamma'\to \Gamma)$, the second argument of the min
function is inverted.

\subsection{Vertex fusion/fission}

\begin{figure}
  \begin{center}
  \qquad
\includegraphics[width=0.5\linewidth]{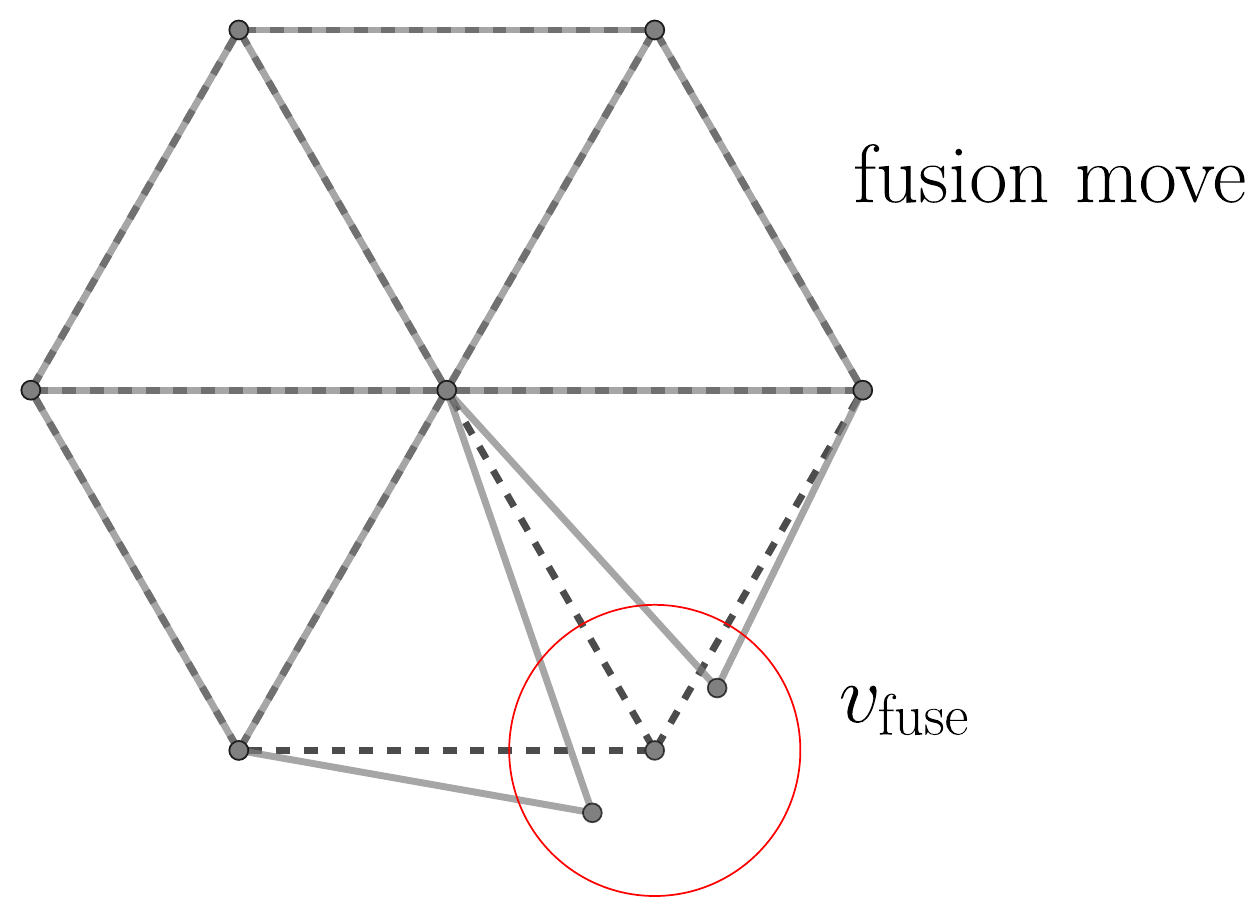}
\end{center}
  \caption{A schematic depiction of the stitching fusion move.}
  \label{fig:fusionmove}
\end{figure}

The insertion moves described above represent events in which shell
monomers arrive from bulk solution and bind to the growing sheet.  A
separate move is required to allow for binding between monomers that
{\em already} belong to the sheet.  In the triangulated sheet
description, such an event fuses one or more existing vertices, so we
term them ``fusion'' moves (and ``fission'' in reverse).  At each
sweep we attempt a fusion move with probability $N_{\rm fuse} k_{\rm
  fuse}^0 \tau$, where $N_{\rm fuse}$ is the number of vertices eligible for fusion
events.  The reverse moves, i.e., fission, are attempted with
probability $N_{\rm fiss} k_{\rm fuse}^0 \tau$, where $N_{\rm fiss}$
is the number of eligible fission events. 
 In each case one of the eligible moves is selected at random and then proposed as detailed below.

Some fusion moves act to close gaps between monomers that already
share a common vertex, as depicted in Fig.~\ref{fig:fusionmove}.  In such a
``stitching'' move, two vertices $i$ and $j$, initially separated by a
distance less than $l_{\rm fuse}$, are merged into a single vertex. The
new vertex is placed
at the midpoint of $i$ and $j$. The generation probability for a
stitching move is
\begin{equation}
  \textrm{gen}(\Gamma\to \Gamma') = N_{\rm fuse}(\Gamma)
  k_{\rm fuse}^0 \tau
  \frac{1}{
    N_\textrm{\rm fuse}(\Gamma)},
\end{equation}
The corresponding change in grand canonical
weight is
\begin{equation}
  \frac{\bar{p}(\Gamma')}{\bar{p}(\Gamma)} = K
  \exp \left( -\beta \Delta {\cal H} \right).
\end{equation}

The reverse of a stitching move splits one surface vertex $i$ into
two.  One of the new vertices is placed at a position randomly
distributed within a spherical volume $v_{\rm fuse}= 4\pi l_{\rm
  fuse}^3/3$ centered on the existing vertex. The other new vertex is
placed opposite the original vertex, such that $i$ coincides with the
midpoint of the new vertices.  The resulting generation probability
\begin{equation}
  \textrm{gen}(\Gamma'\to \Gamma) = N_{\rm fiss}(\Gamma)
  k_{\rm fuse}^0 \tau
  \frac{1}{
    v_{\rm fuse}
    N_\textrm{\rm fiss}(\Gamma)},
\end{equation}
The acceptance probability for a stitching fusion move is thus
\begin{equation}
  \textrm{acc}(\Gamma\to \Gamma') = \min\left[1, \frac{K}{v_{\rm fuse}}
    e^{-\beta \Delta {\cal H}} \right].
\end{equation}
In the acceptance probability for the corresponding fission move,
$\textrm{acc}(\Gamma'\to \Gamma)$, the second argument of the $\min$
function is inverted.

For a shell that is sufficiently large and sufficiently curved, two
different parts of the perimeter can meet. When this happens, binding
of the two parts should be quite facile, since it is not necessary to
await arrival of a new monomer from the dilute solution. Physically,
this union event is just another instance of two shell monomers
binding. Algorithmically, it is more complicated, involving a highly
nonlocal change in vertex connectivity.

We permit such a nonlocal connection move only when two edges at the
sheet's perimeter nearly coincide. Denoting the vertices of one edge
as $i_1$ and $i_2$, and those of the other edge as $j_1$ and $j_2$, we
require that $i_1$ and $j_1$ are separated by a distance less than
$l_{\rm fuse}$ {\em and} that $i_2$ and $j_2$ also reside within the
short distance $l_{\rm fuse}$. One new fused vertex is placed at the
midpoint of $i_1$ and $j_1$; the other is placed at the midpoint
of $i_2$ and $j_2$. The resulting generation probability is
\begin{equation}
  \textrm{gen}(\Gamma\to \Gamma') = N_{\rm fuse}(\Gamma)
  k_{\rm fuse}^0 \tau
  \frac{1}{
    N_\textrm{\rm fuse}(\Gamma)},
\end{equation}
and the change in statistical weight is 
\begin{equation}
  \frac{\bar{p}(\Gamma')}{\bar{p}(\Gamma)} = K^2
  \exp \left( -\beta \Delta {\cal H} \right).
\end{equation}
In a nonlocal fission move, an edge is split in two. The new vertices
are placed as described for stitching fission, giving a generation
probability
\begin{equation}
  \textrm{gen}(\Gamma'\to \Gamma) = N_{\rm fiss}(\Gamma)
  k_{\rm fuse}^0 \tau
  \frac{1}{
    v_{\rm fuse}^2
    N_\textrm{\rm fiss}(\Gamma)}.
\end{equation}
The Metropolis acceptance probability for nonlocal fusion is finally
\begin{equation}
  \textrm{acc}(\Gamma\to \Gamma') = \min\left[1, \frac{K^2}{v_{\rm fuse}^2}
    e^{-\beta \Delta {\cal H}} \right].
\end{equation}
In the acceptance probability for the corresponding nonlocal fission
move, $\textrm{acc}(\Gamma'\to \Gamma)$, the second argument of the
min function is inverted.

Counting fission moves is more involved than counting fusion moves:
For the local stitching scenario,
a vertex $i$ on the perimeter is eligible for fission along an
edge $ij$ if the vertex $j$ is not on the perimeter.
In the case of nonlocal fission, a perimeter vertex $i$ is eligible
to be split
if it has a
neighbor $j$ which is also on the perimeter and if removal of the edge
does not lead to two topologically disconnected domains on the shell.
In practice, we test this condition efficiently using a breadth first
search.

\subsection{Cargo addition/removal}

\begin{figure}
 \begin{center}
 \includegraphics[width=0.6\linewidth]{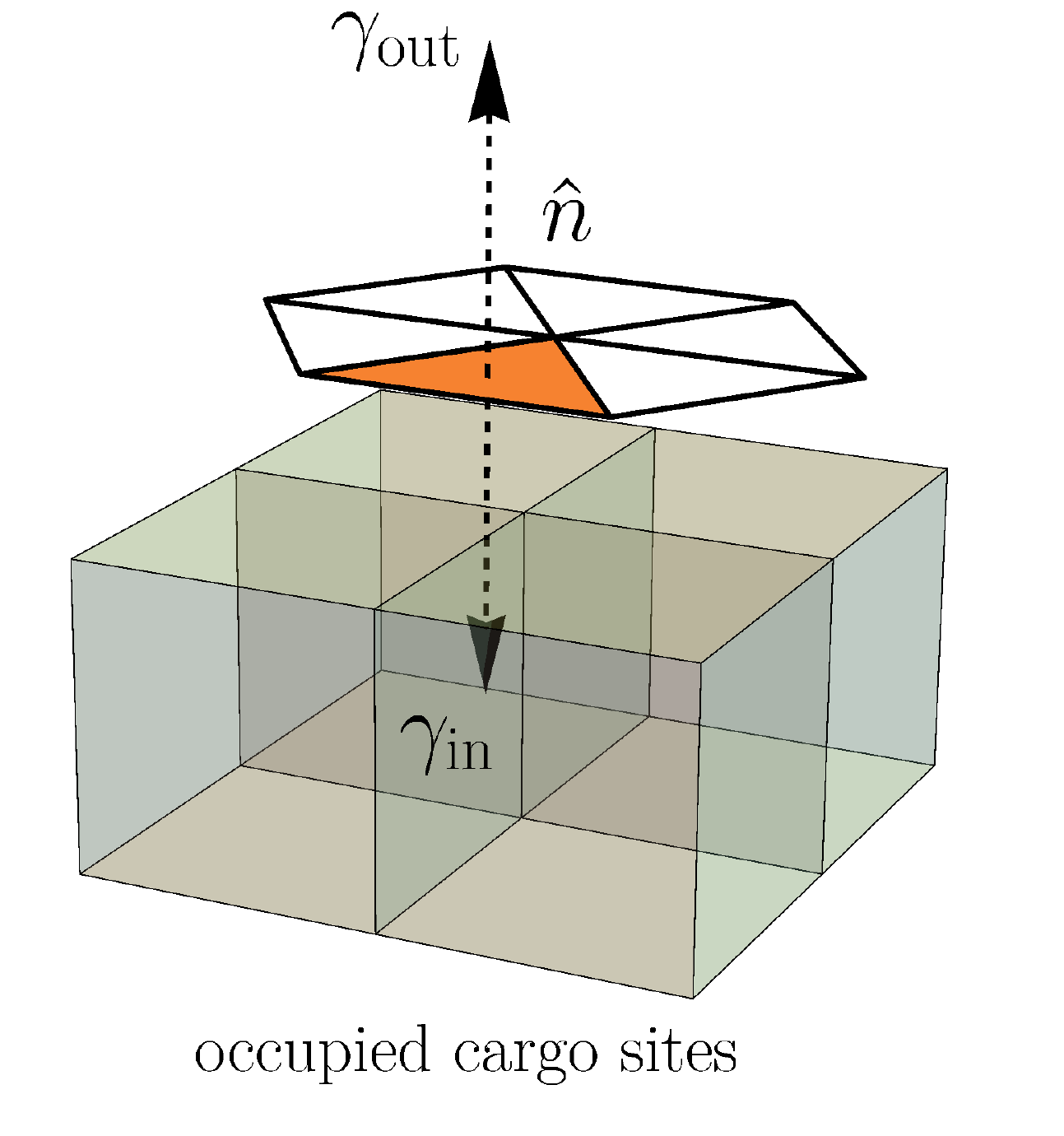}
 \end{center}
 \caption{The cargo molecules, represented as a lattice gas, interact
   with the shell through an interaction $\mathcal{H}_{\rm int}$ that
   is favorable (by an amount $\gamma_\textrm{in}$) if the monomer's
   inward facing normal vector terminates inside a cargo-occupied
   lattice cell. The interaction is unfavorable (by a large amount
   $\gamma_\textrm{out}$) if the outward facing normal vector
   terminates inside a cargo-occupied lattice cell. The latter contribution 
   serves to bias the shell
   fluctuations so that shell monomers are sterically occluded by the
   cargo.}
 \label{fig:interaction}
 \end{figure}

For a given configuration $\{\sigma\}$ the cargo droplet possesses
$N_{\rm boundary}$ sites adjacent to the droplet boundary. This set includes
unoccupied sites with one or more occupied neighbors, where a new
cargo monomer can be accommodated.  It also includes occupied sites
that are incompletely coordinated, where a monomer could be removed
without changing the droplet's connectivity.  During each sweep we
attempt with probability
$N_{\rm boundary} k_{\textrm{c}}^0 \tau$ to change the cargo configuration. We
select one of the boundary sites at random and attempt to change its
state, either inserting or deleting a cargo monomer. The Metropolis
acceptance probability for this procedure is
\begin{equation}
{\rm acc}(\Gamma \to \Gamma') = \textrm{min} \left[1,
  e^{-\beta (\Delta {\cal H}_{\rm cargo} + \Delta {\cal H}_{\rm int})} \right].
\end{equation}

\subsection{Topological defects}
In the absence of cargo, the ground state of a shell protein sheet
features only six-fold coordinated vertices except at the perimeter.
In order to close, however, the shell must develop either vacancies or
else undercoordinated vertices.  Five-fold coordinate sites can arise
as a consequence of insertion or fusion moves, and they play an
important role in encapsulation.

The Monte Carlo moves we have described could also produce
overcoordinated vertices. Because there is no evidence for seven-fold
(or higher) symmetric protein assemblies in the carboxysome
literature, we prohibit these defects from forming, with a single
exception.  A seven-fold defect is allowed to result from fusion that
causes two fivefold defects to share a neighbor.  This arrangement can
be viewed as accommodating a vacancy in the monomer sheet, rather than
creating a heptameric arrangement of shell monomers.  The large
favorable energy for such a fusion event makes the distinction between
vacancy and excess coordination thermodynamically unimportant.

\subsection{Constraints and parameterization}
The constraints on connectivity we impose in these dynamics -- there
is exactly one connected cargo droplet and exactly one connected sheet
-- are artificial.  This is done as a matter of convenience, with
substantial computational savings.  The constraint should not
significantly affect assembly dynamics provided that nucleation of
cargo and shell are both slow compared to their growth.  The physical
conditions we examine fall in this regime.

The set of parameters governing our molecular simulations are listed
in Tables~\ref{table:cargoparams} and~\ref{table:shellparams}.
One key feature is that elastic relaxation is very fast
relative to monomer addition, i.e., $k_{\textrm{s}}^0 \tau \ll 1$ and $k_{\textrm{c}}^0 \tau
\ll 1$. Another is that shell growth is highly unfavorable in the
absence of cargo, as guaranteed by small values of $z_{\textrm{s}} K^2 v_{\rm
  add}$ and $z_{\textrm{s}} K^3$. The potent attraction between shell and cargo
makes simple shell monomer addition feasible but reversible when cargo
is present on the interior side of the new monomer. Wedge insertion
next to cargo is even more favorable and is typically accepted with
high probability.

In some cases, two disconnected edges can come together so as to
enclose a triangular domain, but one which was not explicitly occupied
by a monomer before the fusion event. In the triangulated sheet
representation, the newly enclosed domain is assumed to be occupied by
a shell monomer. Such a move therefore implicitly entails the arrival
and binding of a new monomer. Because this binding is strongly downhill in
free energy for the physical conditions of interest, we treat it
as inevitable and irreversible, and we do not represent it explicitly.

\section{Phenomenological free energy}

\subsection{Macroscopic equilibrium}

In order to assess the equilibrium arrangements of cargo and shell
species, we view their assembly as a chemical transformation, in which $N_{\textrm{s}}$ shell capsomers $S$ combine with $N_{\textrm{c}}$ cargo subunits $C$ to form a complex $A_{N_{\textrm{s}},N_{\textrm{c}}}$, which might or might not be a completely closed microcompartment.
This process can be represented as a simple kinetic scheme
\begin{equation}
  N_{\textrm{s}} S + N_{\textrm{c}} C \rightleftharpoons A_{N_{\textrm{s}},N_{\textrm{c}}}.
\end{equation}

At thermodynamic equilibrium, stability criteria require that the chemical potentials $\mu$ associated with the constituent materials are balanced.
\begin{equation}
N_{\textrm{s}} \mu_{\textrm{s}} + N_{\textrm{c}} \mu_{\textrm{c}}  = \mu_{N_{\textrm{s}},N_{\textrm{c}}}.
\end{equation}
Assuming these particles are present at very dilute concentration, the chemical potential of each species $j$ can be written as a function of its density, the temperature $T$, and a density-independent standard state chemical potential $\mu^{(0)}$,
\begin{equation}
\mu_j(T) = \mu_j^{(0)}(T) + T \ln \frac{\rho_j}{\rho^{(0)}}.
\end{equation}
Note that Boltzmann's constant has been absorbed into the unit of temperature, i.e., $k_{\textrm{B}} = 1$.

Combining the relationships above yields a law of mass action,
\begin{equation}
  \rho_{N_{\textrm{s}},N_{\textrm{c}}} = \rho^{(0)} \exp \left[ - \beta \Delta\mu_{N_{\textrm{s}},N_{\textrm{c}}}^{(0)} - N_{\textrm{s}} T \ln \frac{\rho_{\textrm{s}}}{\rho^{(0)}} -  N_{\textrm{c}} T \ln \frac{\rho_{\textrm{c}}}{\rho^{(0)}}\right].
  \label{equ:massaction}
\end{equation}
The equilibrium constant relating densities of unassembled and assembled species is expressed here in terms of the reversible work required to assemble the components into a filled shell at standard state concentration,
\begin{equation}
  \Delta \mu^{(0)}_{N_{\textrm{s}},N_{\textrm{c}}} = \mu_{N_{\textrm{s}},N_{\textrm{c}}}^{(0)} - N_{\textrm{s}} \mu_{\textrm{s}}^{(0)} - N_{\textrm{c}} \mu_{\textrm{c}}^{(0)}. 
\end{equation}
The simple relationship (\ref{equ:massaction}) among particle densities allows us to investigate the equilibrium distribution of encapsulated structures. 
In particular, we ask at which value $(N_{\textrm{s}}, N_{\textrm{c}})$ the density $\rho_{N_{\textrm{s}},N_{\textrm{c}}}$ is maximal. 

Equivalently, we can find a minimum of the free energy
\begin{equation}
  G(N_{\textrm{s}},N_{\textrm{c}}) = \Delta \mu^{(0)}_{N_{\textrm{s}},N_{\textrm{c}}} - N_{\textrm{s}} T \ln \frac{\rho_{\textrm{s}}}{\rho^{(0)}} -  N_{\textrm{c}} T \ln \frac{\rho_{\textrm{c}}}{\rho^{(0)}}.
\end{equation}
While an exact expression for $G$ is not analytically tractable for
our molecular model, we can judge from the energy scales involved what
type of minima might exist.  In particular, it is clear that the
shell-shell and cargo-shell interactions have an energy scale
proportional to $N_{\textrm{s}}$.
while the cargo-cargo interactions have an energy
proportional to $N_{\textrm{c}}.$ Using these intuitive scaling arguments, we can
postulate a simple expression for the free energy
\begin{equation}
 G(N_{\textrm{s}},N_{\textrm{c}}) =  a N_{\textrm{s}} + b N_{\textrm{c}} + E_\textrm{el}(N_{\textrm{s}},N_{\textrm{c}}),
 \end{equation}
 where the constants $a$ and $b$ are unknown functions of the densities, temperature, and binding free energies.
 
 The elastic energy, in the case of an idealized spherical geometry, does not scale with the size of the shell; it is a constant $E_\textrm{el}(N_{\textrm{s}}, N_{\textrm{c}}) = 4\pi\kappa.$
 In the more general setting of nonspherical geometries, we can approximate the free energy as
 \begin{equation}
  G(N_{\textrm{s}},N_{\textrm{c}}) =  a \zeta N_{\textrm{c}}^{2/3} + b N_{\textrm{c}} + \textrm{const.} + E_\textrm{el}(\zeta),
 \end{equation}
       where $\zeta \equiv
     N_{\textrm{s}}N_{\textrm{c}}^{-2/3}$ is a parameter that quantifies the asphericity of the
     structure.  Note that $\zeta$ is not the same quantity used to
     characterize the sphericity in Sec. S7.  A detailed analysis of
     the extrema of the free energy functional gives insight into the
     types of shapes that are thermodynamically preferred.  The free
     energy has a minimum value $\zeta_\textrm{min}=O(1)$ in which the
     structural is spherical.  Because the elastic energy
     $E_\textrm{el}$ is an increasing function of
     $\zeta$, the minimum
     of the free energy will have $\zeta = \zeta_\textrm{min}$ for a
     fixed $N_{\textrm{c}}$, so long as $a>0.$ As a result, global minima of the
     free energy can only occur at $N_{\textrm{c}}=0$ or $N_{\textrm{c}}=\infty$, depending on
     the sign of $b.$ Mathematically, negative values of $a$ could
     produce other minima, but such a negative surface tension would result
     in highly distended shapes, which are not physically relevant for
     our model.  A similar scaling argument can be made for idealized
     icosahedral structures.  
       This accounting suggests that the metastable
     microcompartment structures seen in biological systems
     represent a departure from equilibrium statistics.

\subsection{Thermodynamic potential for compartment size and shell coverage}
A more detailed phenomenological free energy functional can be
constructed by assuming that cargo-shell assemblies adopt the simple
geometry suggested by the analysis above.  Specifically, cargo
molecules are assumed to form a compact spherical globule, and the
shell proteins are assumed to coat the globule as a spherical cap.  At
the level resolved here, icosahedral structures would be governed by a
very similar potential.

It is convenient to write this functional in terms of the radius of the spherical cargo droplet, $R$, and the angle subtended by the spherical cap $\theta$.
We let $f_{\textrm{c}}$ denote the free energy per unit cargo and $\gamma_{\textrm{c}}$ the surface tension associated with the condensed cargo droplet. 
Similarly, we define $f_{\textrm{s}}$ and $\gamma_{\textrm{s}}$ as the free energy per unit shell and the line tension around the growth edge of the shell. 
The variables $\rho$ and $\nu$ denote the density and surface density of the cargo and the shell proteins, respectively.

The shell-cargo interaction energy is $-\epsilon$ per monomer. 
The elastic energy is determined by a bending rigidity $\kappa$.
Finally, the activity of the cargo monomers is $z_{\textrm{c}}$ and the activity of the shell monomers is $z_{\textrm{s}}$.
Putting together all the terms, we have 
\begin{equation}
  F(R, \theta) = \frac{4}{3} \pi R^3 \rho (f_{\textrm{c}} - T \ln z_{\textrm{c}}) + 2\pi R^2 (1-\cos\theta) \nu (f_{\textrm{s}} - T\ln z_{\textrm{s}} - \epsilon) + 4\pi R^2 \gamma_{\textrm{c}} + 2\pi R \sin\theta \gamma_{\textrm{s}} + 2\pi \kappa (1-\cos \theta).
  \label{eq:phenom}
\end{equation}
The fully assembled microcompartment is stable when $\Delta F_\textrm{close}(R) \equiv F(R,\pi) - F(R,0)$ becomes negative.
If the rate of cargo growth is much slower than the rate of shell growth, the expected size of the assembly is
\begin{equation} 
R^* = \sqrt{\kappa/|\nu(f_{\textrm{s}}-T\ln z_{\textrm{s}} - \epsilon)|}
\end{equation}
The average size of an encapsulated structure thus depends strongly on the bending rigidity $\kappa$, suggesting one way to tune microcompartment
assembly; controlling monomers' detailed physical properties, however,
is likely a difficult engineering problem in practice.
The shell activity $z_{\textrm{s}}$ is a much more pragmatic control parameter
for this purpose.

\begin{figure}[ht!]
 \begin{center}
 \includegraphics[width=0.85\linewidth]{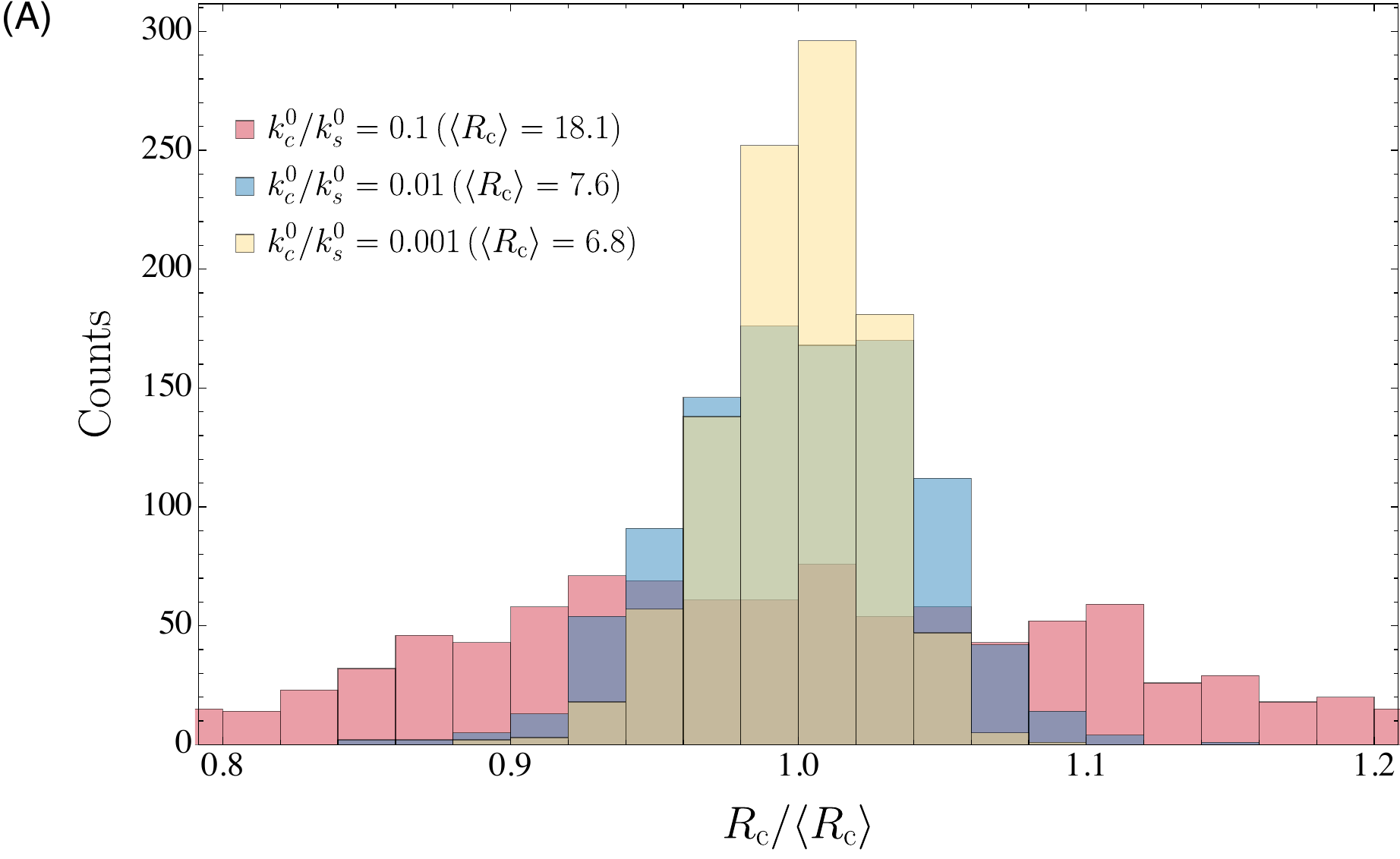}
 \includegraphics[width=0.85\linewidth]{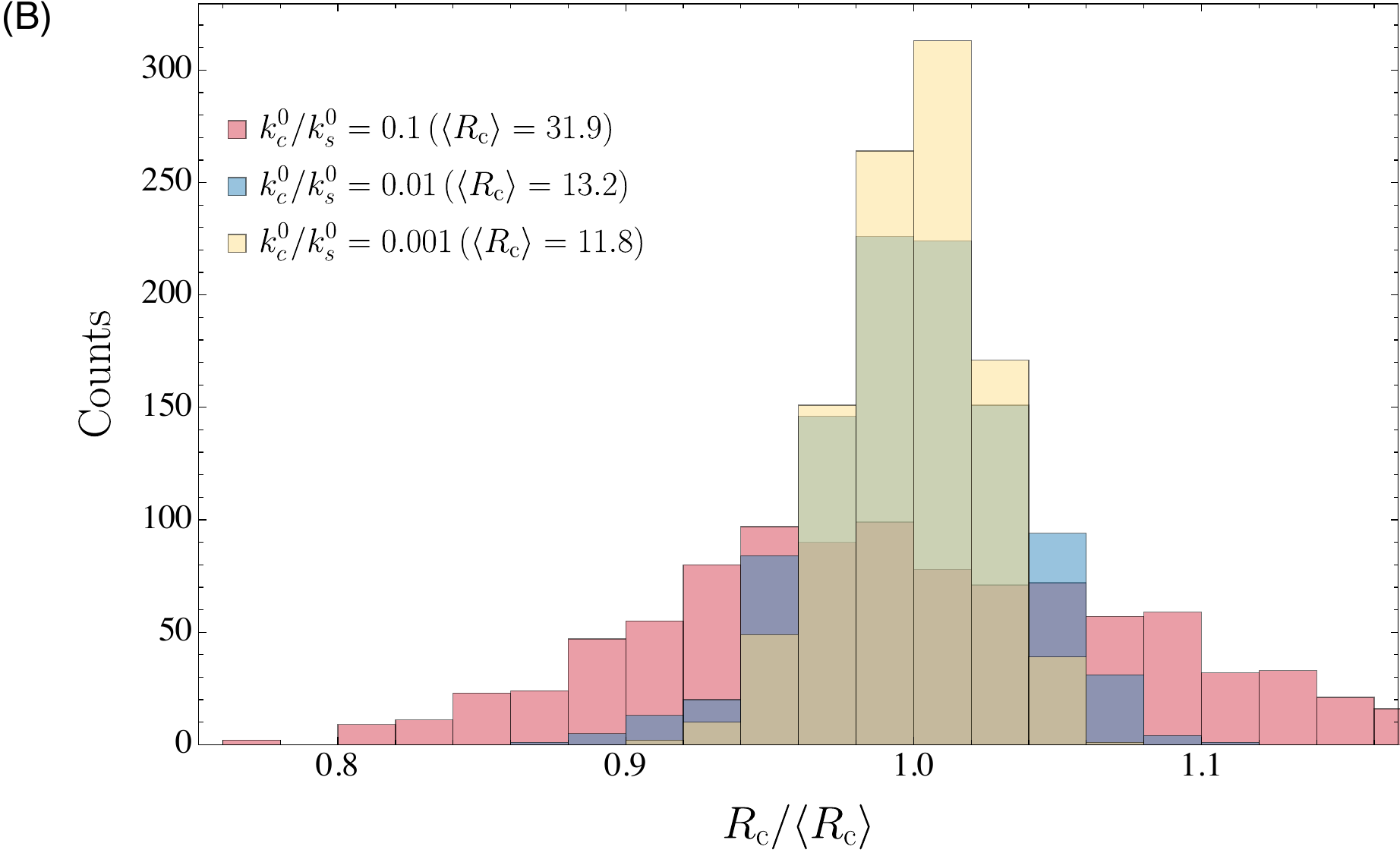}
 \end{center}
 \caption{Histograms of the radii of the encapsulated structures
   computed from kinetic Monte Carlo simulations. (A)
   $\kappa=100$. (B) $\kappa=300.$
   Data was
   collected from $10^4$ independent KMC simulations.
 Cargo radii are given relative to the length scale $\ell= (2\pi\nu)^{-1/2}$.}
 \label{fig:rhists}
 \end{figure}

Both the shell and the cargo may need to overcome nucleation barriers before they begin growing spontaneously.
In practice, this means that the expected size of the completely assembled structure will depend on the height and location of these barriers.
The free energy $F$ can be written in a form that highlights the connection to classical nucleation theory as
\begin{equation}
  F(R, \theta) = a_\textrm{c} \left[ \left( \frac{R}{R_\textrm{c}} \right)^2 - \left( \frac{R}{R_\textrm{c}} \right)^3 \right] + a_\textrm{s} \left[ \left\{ \left(\frac{R^*}{R_\textrm{s}} \right)^2 - \left(\frac{R}{R_\textrm{s}} \right)^2 \right\}\left(1-\cos \theta\right) + \frac{R}{R_\textrm{s}} \sin \theta \right],
\end{equation}
with 
\begin{eqnarray}
a_{\textrm{s}} &= \frac{2\pi \gamma_{\textrm{s}}^2}{\nu (T\ln z_{\textrm{s}}+\epsilon -f_{\textrm{s}})} \\
R_{\textrm{s}} &= \frac{\gamma_{\textrm{s}}}{\nu (T\ln z_{\textrm{s}} + \epsilon -f_{\textrm{s}})} \\
a_{\textrm{c}} &= \frac{36 \pi \gamma_{\textrm{c}}^3}{\rho^2 (f_{\textrm{c}} - T\ln z_{\textrm{c}})^2} \\
R_{\textrm{c}} &= \frac{3 \gamma_{\textrm{c}}}{\rho (T\ln z_{\textrm{c}} - f_{\textrm{c}})} \\
R^* &= \sqrt{2\pi \kappa} R_{\textrm{s}}.
\end{eqnarray}
This expression highlights the characteristic sizes of the critical nuclei and their interconnected dependence on the microscopic parameters.
In particular, it should be noted that for the parameters that we are exploring the cargo has essentially no barrier to nucleation, whereas the shell has a significant one at small radius.
However, as the radius increases, the shell can grow with greater facility and encapsulation becomes kinetically accessible.
Fig.~\ref{fig:rhists} emphasizes that, even for large average radii, the distribution of sizes can be made tight by decreasing the rate of cargo addition.

In simulations of the minimalist model, we adopt a unit of length
$\ell = (2\pi\nu)^{-1/2}$, choose a cargo density $\rho =
3/(4\pi\ell^3)$, and set $L_0=\ell$, simplifying many of the above
expressions. In particular, with these choices the monomer populations
can be written transparently in terms of geometric variables as
$N_\textrm{s} = R^2 (1-\cos\theta)$ and $N_\textrm{c}=R^3$.

 \begin{figure}[ht!]
 \begin{center}
 \includegraphics[width=0.7\linewidth]{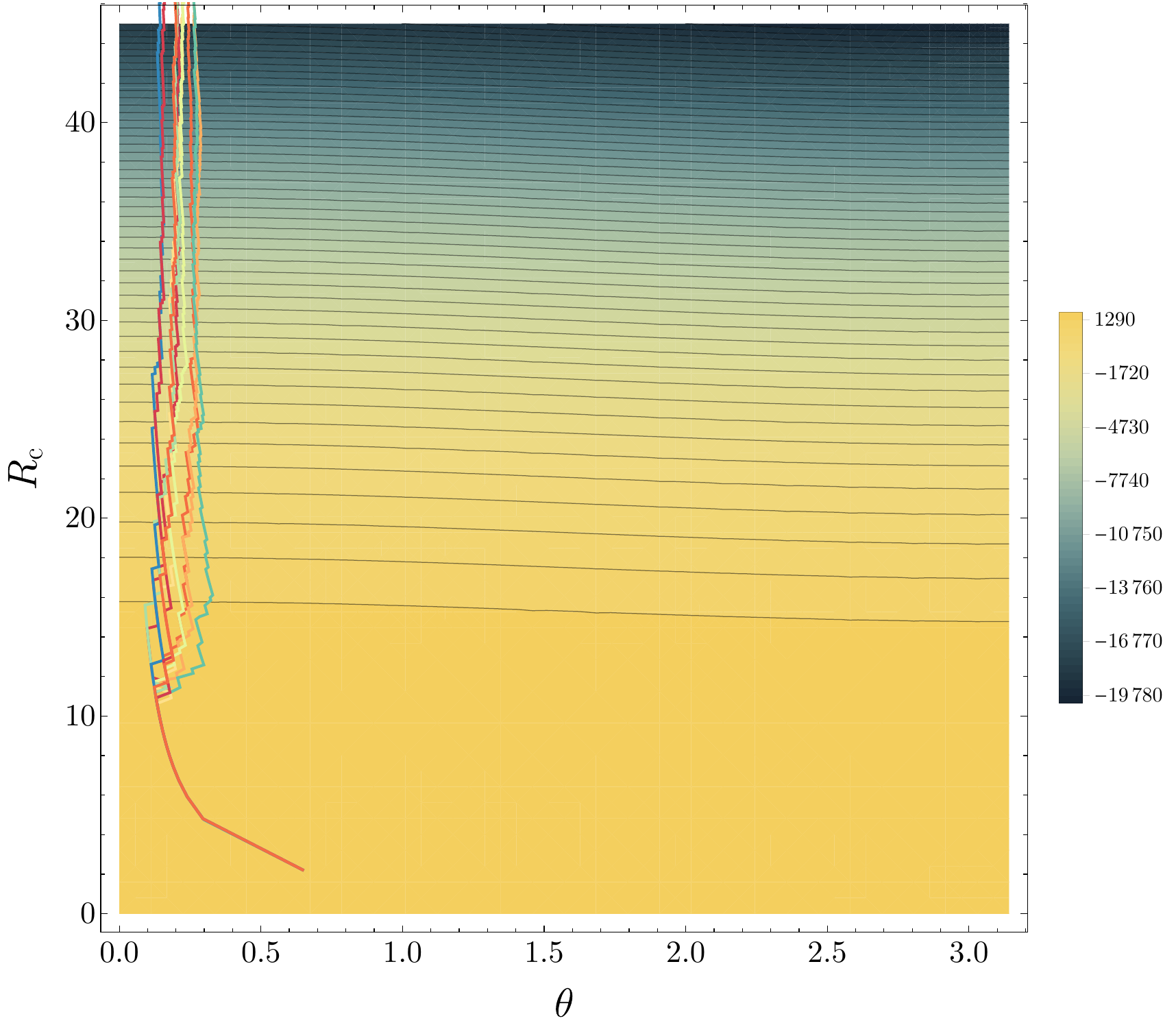}
 \end{center}
 \caption{Trajectories of the minimal model illustrating runaway
   growth. The cargo and shell arrival rates in this case are equal,
   $k_{\textrm{c}}^0/k_{\textrm{s}}^0 = 1$. The trajectories do not
   encapsulate after $10^6$ steps of KMC dynamics.  Cargo radius is
   given relative to the length scale $\ell= (2\pi\nu)^{-1/2}$. }
 \label{fig:runaway}
 \end{figure}

\section{Simulations of the molecular model}

All simulations were performed using an implementation of the algorithm described in Section~\ref{sec:mcmc} which we have open sourced. It is available on GitHub: \hyperref[https://rotskoff.github.io/shell-assembly/]{rotskoff.github.io/shell-assembly/}.
The parameters used for all simulations are described in Tables S2 and S3.

\begin{table}
\caption{Simulation parameters for the FCC lattice gas cargo.
All energies are given in units of $k_{\rm B}T$.} 
\begin{center}
  \begin{tabular}{|l|c|c|}
\hline
  $\epsilon_\textrm{C}$ & 1 & cargo binding energy scale \\ \hline
  $\mu_\textrm{C}$ & -5 & cargo chemical potential \\ \hline
  $\gamma_\textrm{in}$ & 5 & cargo-shell interaction energy (inward normal) \\ \hline
  $\gamma_\textrm{out}$ & 100 & cargo-shell interaction energy (outward normal)\\ \hline
    $k_{\textrm{c}}^0\tau$ & 0.000075  & rate of cargo insertion attempts\\ \hline
\end{tabular}
\end{center}
 \label{table:cargoparams}
\end{table}

\begin{table}

\caption{Simulation parameters for the elastic shell model.
  All energies are given in units of $k_{\rm B}T$, and lengths in
  units of $\ell_0$.}
\begin{center}
\begin{tabular}{|l|c|c|}
\hline
  $\epsilon$ & 500. & stretching energy scale \\ \hline
  $\kappa$ & 12.5 & bending energy scale \\ \hline
  $l_0$ & 1. & equilibrium shell bond length\\ \hline
  $l_\textrm{max}$ & 1.5 $l_0$ & shell bond length cutoff \\ \hline
  $l_\textrm{min}$ & 0.5 $l_0$ & shell bond length cutoff \\ \hline
  $l_\textrm{fuse}$ & 0.5 $l_0$ & distance cutoff for fusion attempts\\ \hline
  $l_\textrm{add}$ & 0.5 $l_0$ & radius for fusion proposal\\ \hline
  $\ln K$ & 7 & binding affinity scale parameter\\ \hline
  $\ln z$ & -16.5 & activity parameter \\ \hline
  $k_{\textrm{s}}^0\tau$ & 0.0075 & rate of shell insertion attempts\\ \hline
\end{tabular}
\end{center}
\label{table:shellparams}
\end{table}

All simulations were initialized with a hexamer of shell material which we did not allow to be removed.
This ensures that nucleation of the shell can occur. 
Cargo is initiated by populating all lattice sites within three lattice
spacings of the origin (corresponding to a total number 429 of cargo monomers).

The simulation completes when there are no remaining surface edges on the shell structure. 
At this point, growth moves are no longer proposed and the structure is equilibrated for $10^5$ MC sweeps before exiting. 
An animation of a trajectory of the microscopic model is shown in Movie S1.

Simulations were visualized using the VMD software package~\cite{vmd}. 
In order to clearly show the shell structure, only occupied lattice sites that are above a distance cutoff of 1.5$\l_0$ are shown in our visualizations. 
The cargo is depicted either as density isosurfaces or spheres centered at the lattice sites. 
In the Movie S1, the shell is depicted as dynamic bonds with a distance cutoff of $1.5l_0,$ which leads to unconnected vertices appearing bonded in some frames.

\section{Free energy calculations from the molecular model}

\begin{figure}[ht!]
   \begin{center}
     \includegraphics[width=0.45\linewidth]{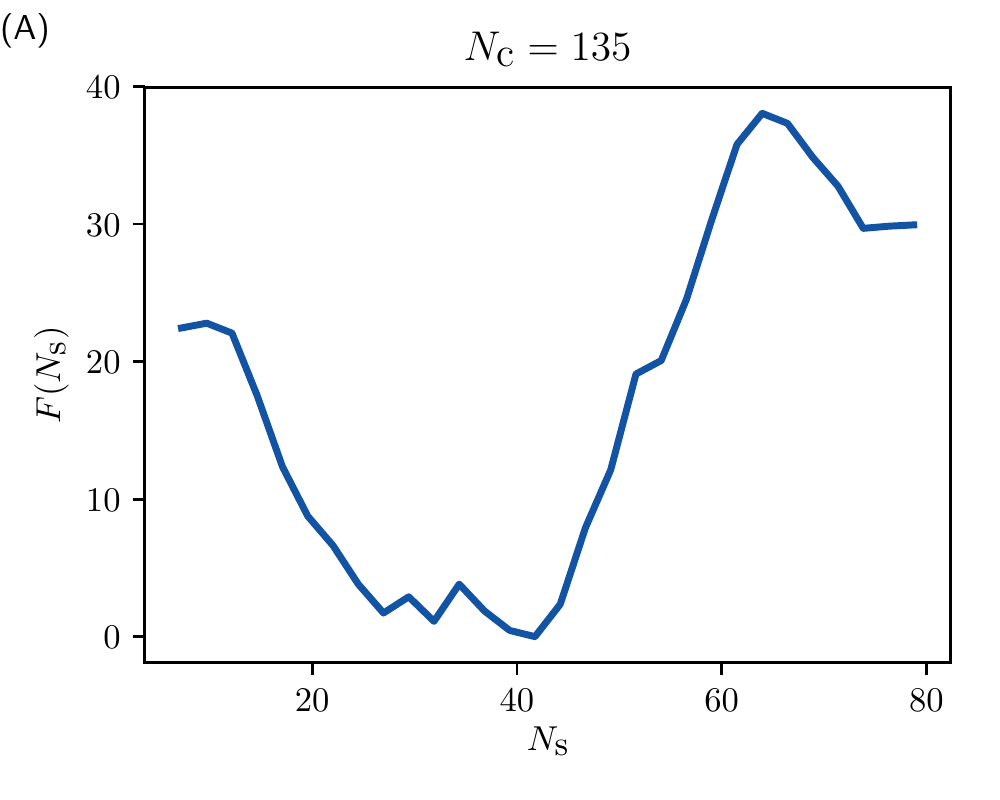}
     \includegraphics[width=0.45\linewidth]{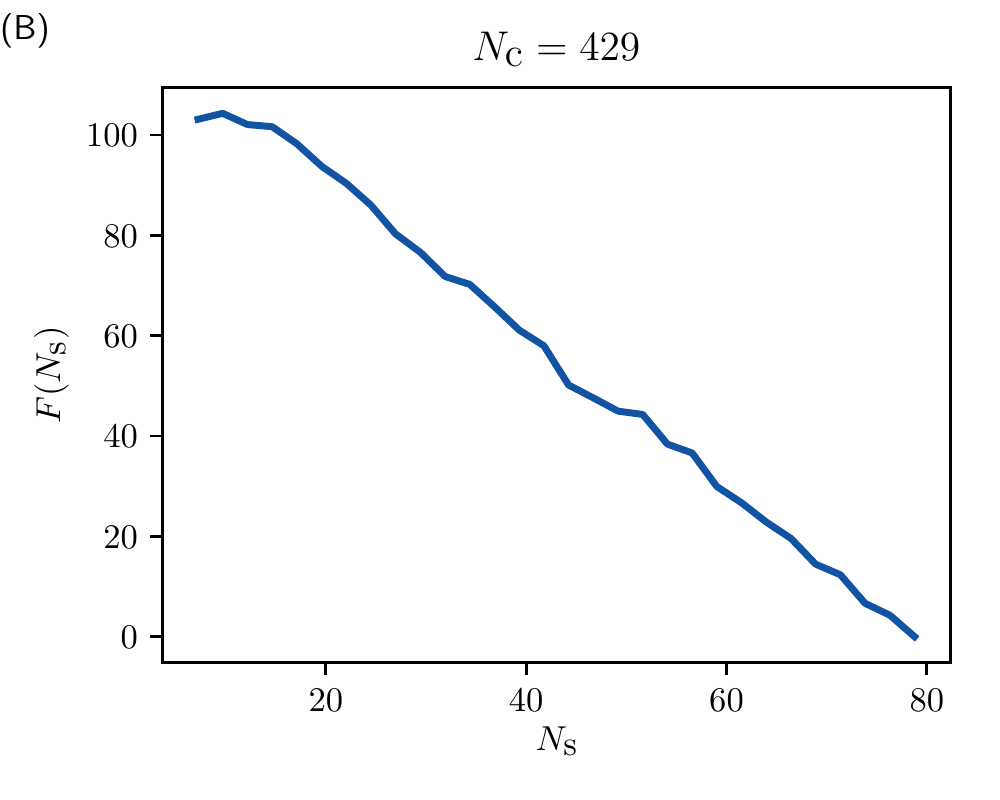}
     \includegraphics[width=0.45\linewidth]{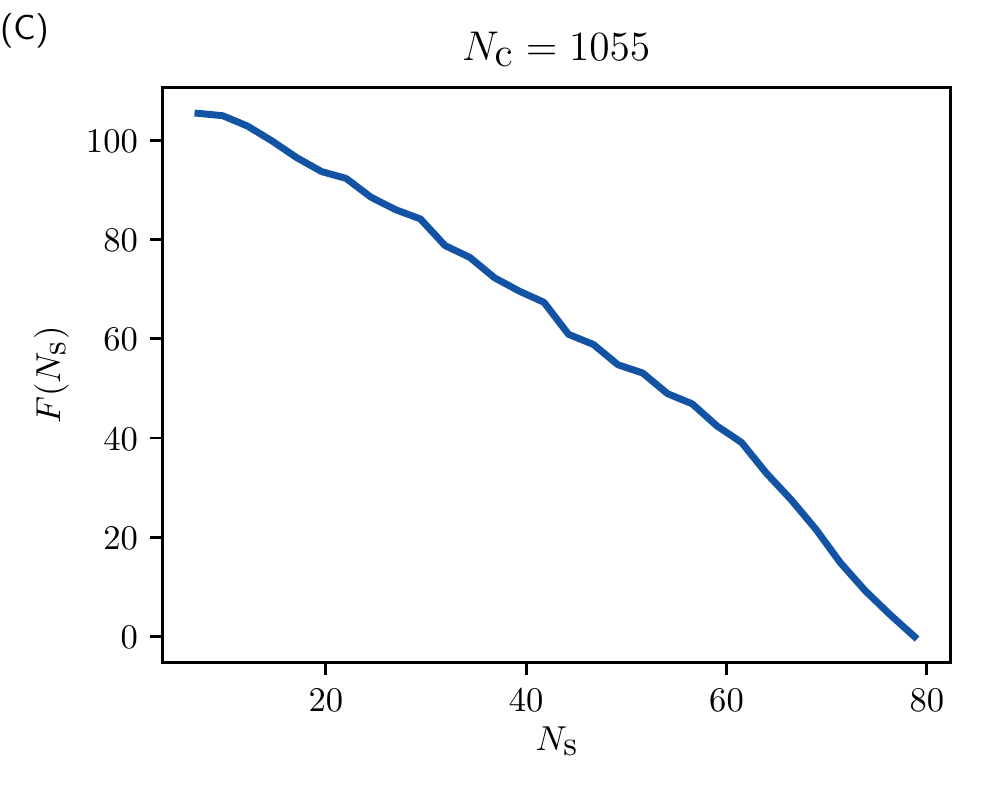}
          \includegraphics[width=0.45\linewidth]{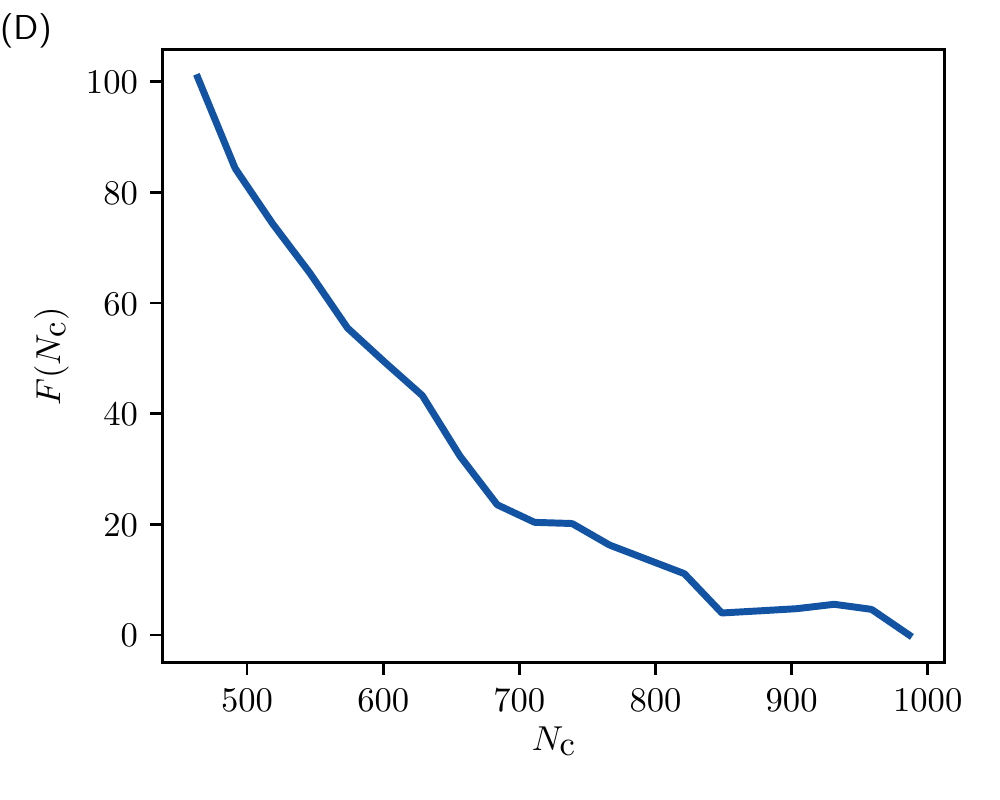}
   \end{center}
   \caption{Umbrella sampling calculations of the molecular model.
     Free energies $F$ are given in units of $k_{\rm B}T$. In
     panels A-C we show free energy
     as a function of shell monomer population, with the cargo size
     fixed at the indicated value. In panel D we show free energy
   as a function of cargo population, with the shell size unconstrained.}
     \label{fig:umb}
\end{figure}

We performed a series of umbrella sampling calculations to compare the
thermodynamics of our molecular model with the minimalist model.
First, we considered a small cargo droplet $N_{\rm c}=135,$
corresponding to a sphere with diameter $4 l_0.$ As shown in
Fig.~\ref{fig:umb}, the cost of deforming the shell makes
encapsulation uphill in free energy.  In the case of a droplet with
$N_{\rm c}=429,$ a sphere of diameter $6 l_0,$ which is above the
nucleation barrier for cargo growth, we find that there is a modest
barrier to the initial growth which becomes downhill as $N_{\rm s}$
increases.  Finally, we calculated the free energy for $N_{\rm
  c}=1055$, corresponding to sphere with diameter $10 l_0.$ In
addition to the free energy calculations performed for shell growth,
we also performed a free energy calculation to show that cargo growth
is downhill for the regime of interest.

All umbrella sampling calculations for $N_{\rm s}$ were performed with a harmonic bias energies with force constant $k=1.0,$ and centers separated by two monomers. 
The cargo umbrella sampling was performed with $k=1.0$ and windows separated by $5$ cargo monomers.

\section{Dynamical flows for the minimalist model}

 \begin{figure}[ht!]
 \begin{center}
 \includegraphics[width=0.4\linewidth]{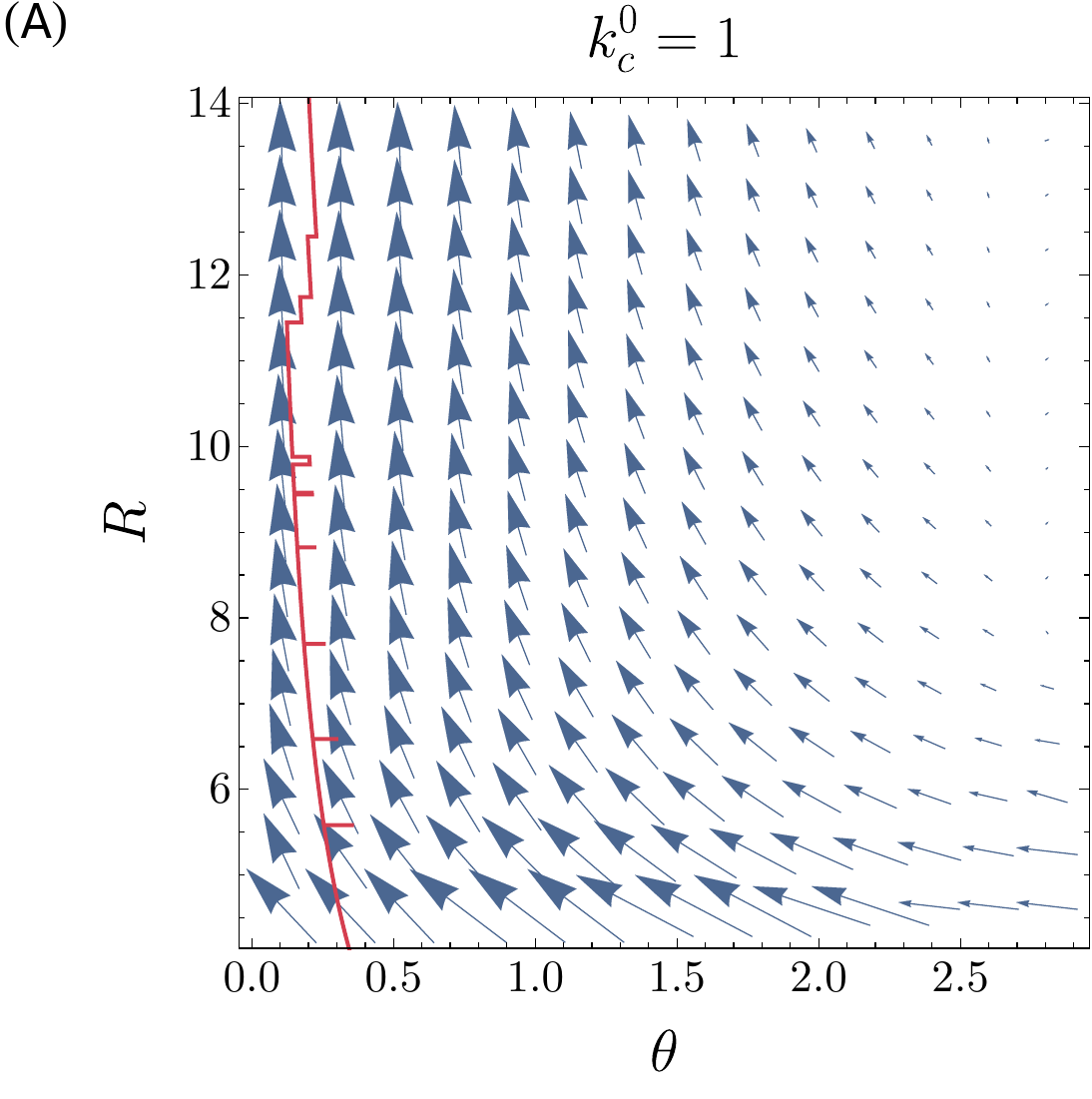}
 \includegraphics[width=0.4\linewidth]{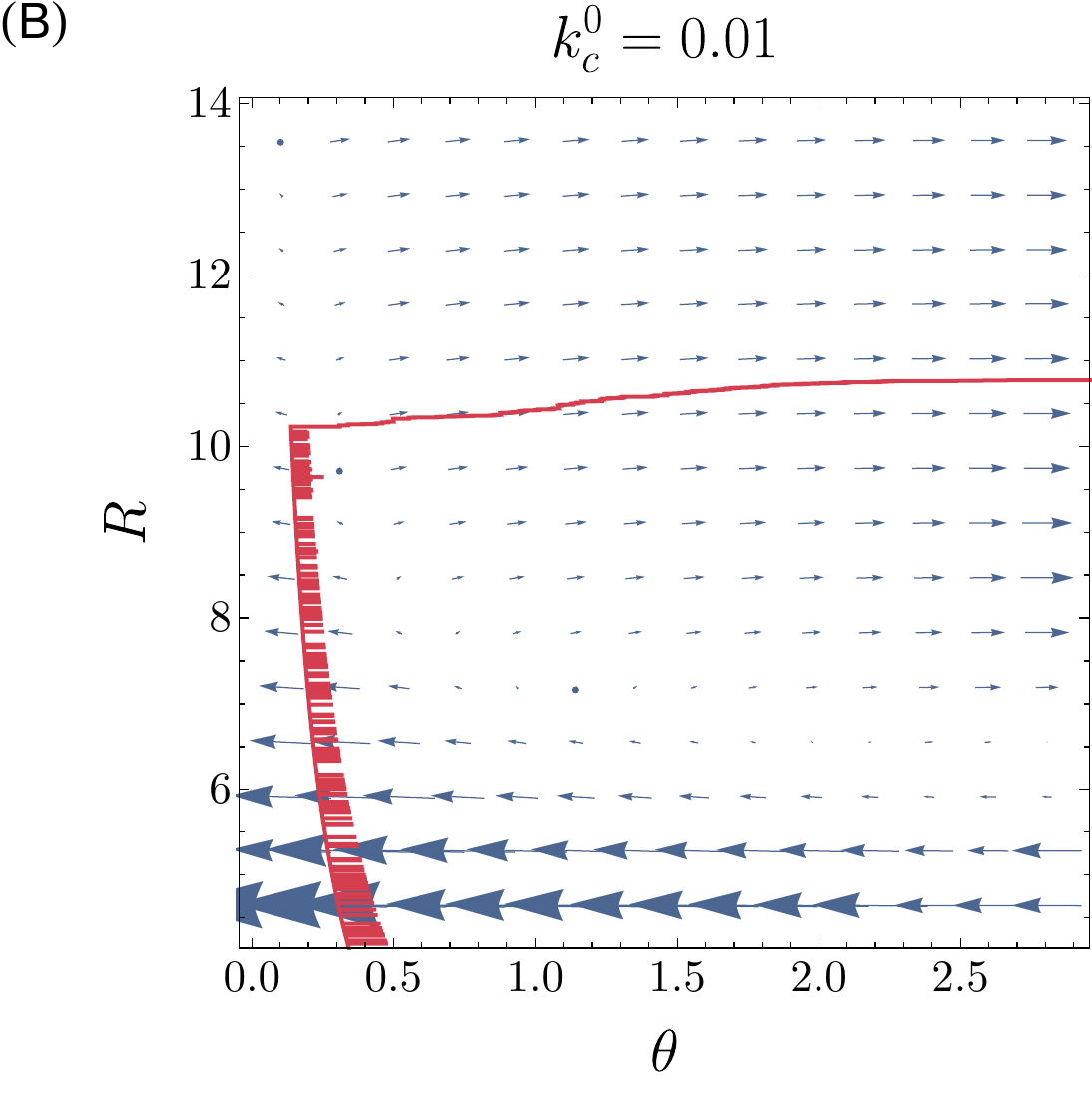}
 \end{center}
 \caption{Flow vectors $\omega = (\dot{\theta},\dot{R})$ are plotted with representative trajectories. The size of the vector is proportional to its magnitude. In both cases, $\kappa=200$ and the shell arrival rate $k_{\textrm{s}}^0=1$. (A) A representative trajectory with $k_{\textrm{c}}^0 = 1$
   is shown in red below the vector field. (B) A representative trajectory with $k_{\textrm{c}}^0=0.01$ is shown in red below the vector field. }
 \label{fig:flows}
 \end{figure}

In very simple models, the routes of typical trajectories can be
inferred directly from the underlying free energy surface.  Kinetic
inference is not so straightforward for our minimalist model.
Dynamical flows are complicated by geometric factors,
e.g., the fact that cargo addition rates are proportional to the
number of available surface sites, which changes during the course of
assembly.  The free energy $F(R,\theta)$ in this case constrains
ratios of rates for each forward and reverse move, but does not give a
good sense for the directions along which the system typically
evolves.

In this section we develop a means to visualize dynamical tendencies for cargo droplet growth and shell envelopment in our minimalist model. 
Let ${\bf \omega}(\theta,R)$ be a vector field with components $\omega_R=\dot{R}$ and $\omega_\theta=\dot{\theta}$ along $R$ and $\theta$ that reflect the typical rate of change along these coordinates at each point in the $\theta,R$-plane. 
First, we differentiate the relationships
\begin{align}
  R(N_{\textrm{c}}) &= \left(\frac{3}{4\pi}(L_0 \rho)^{-1} N_{\textrm{c}}\right)^{1/3} \\
  \theta(N_{\textrm{s}},N_{\textrm{c}}) &= \cos^{-1}\left(1-\frac{N_{\textrm{s}}l_0}{2 \pi \nu R(N_{\textrm{c}})^2}\right),
\end{align}
to express the time derivatives $\dot{R}$ and $\dot{\theta}$ in terms of $R$, $\theta$, $\dot{N}_{\textrm{s}}$ and $\dot{N}_{\textrm{c}}$, where dots represent differentiation with respect to time.
We then estimate average rates of change in $N_{\textrm{s}}$ and $N_{\textrm{c}}$ through the effective drift in these populations,
\begin{align}
\dot{N}_{\textrm{s}} &= -k_{\textrm{s}}^+(N_{\textrm{s}}, N_{\textrm{c}}) \pd{G}{N_{\textrm{s}}}\\ 
\dot{N}_{\textrm{c}} &= -k_{\textrm{c}}^+(N_{\textrm{s}}, N_{\textrm{c}}) \pd{G}{N_{\textrm{c}}},
\end{align}
which can in turn be written in terms of $R$ and $\theta$. 
Plotting ${\bf \omega}$ as a function of $R$ and $\theta$ yields a dynamical flow field that reveals the direction in which trajectories typically evolve from each point $(\theta,R)$.

The dynamical flow fields shown in Fig.~\ref{fig:flows} highlight the importance of monomer arrival rates, particularly for the likelihood of shell closure. 
Panels (A) and (B) correspond to systems with the same free energy $F(R,\theta)$, but with different relative arrival rates, $k_{\textrm{c}}^0$, for cargo.  

For very fast cargo arrival, the nucleation of sufficient shell material to encapsulate the rapidly expanding surface of the cargo is kinetically impeded. 
The field shown in Fig.~\ref{fig:flows} (A) corresponds to this case of ``runaway'' growth (see also, Fig.~\ref{fig:runaway}), with ${\bf \omega}$ pointing primarily in the direction of increasing $R$ throughout much of $(\theta,R)$ space.
In a typical trajectory, the radius of the cargo increases rapidly and the angle $\theta$ cannot grow to fully encapsulate the surface.

By contrast, when the shell arrival rate greatly exceeds that of cargo, the field points more strongly along increasing $\theta$, promoting shell closure. 
As soon as the barrier to nucleating a shell disappears, the average dynamical flow leads trajectories rapidly and unequivocally towards encapsulation, as shown in Fig.~\ref{fig:flows} (B). 
This representation of dynamical biases does not capture the possibility of barrier crossing, i.e., the spontaneous nucleation of a substantial shell even when it requires activation. 
In these systems, however, this nucleation barrier becomes comparable to $k_\textrm{B}T$, and therefore easily navigated, not far from the point where shell growth becomes thermodynamically downhill.

The location of the barrier $\theta^\ddagger(R)$, for a given droplet
radius $R$, can be determined by finding the maximum of
$\pd{F(R,\theta)}{\theta}$ in the range $[0,\pi].$ Using the resulting
expression,
\begin{equation}
  \theta^\ddagger(R) = \tan^{-1}\left( \frac{\frac{R}{R_{\textrm{s}}}}{\left(\frac{R}{R_{\textrm{s}}}\right)^2 - \left( \frac{R^*}{R_{\textrm{s}}} \right)^2} \right),
  \label{eq:thetabar}
\end{equation}
we can estimate the radius at which the barrier to shell growth
becomes comparable to the thermal energy.  We judge the height of the
barrier,
\begin{equation}
\Delta F^\ddagger(R) = F(R,\theta^\ddagger) - F(R,\theta_0),
\end{equation}
with reference to a typical value $\theta_0=0.2$ of the spherical cap
angle prior to nucleation of a steadily growing shell. This activation
free energy becomes comparable to thermal energy at a radius $R^\ddagger$
for which
\begin{equation}
  \Delta F^\ddagger(R^\ddagger) = k_\textrm{B} T.
\end{equation}
The critical radius $R^\ddagger$ compares favorably with the point at
which shell nucleation occurs in kinetic Monte Carlo simulations, as shown in Fig. 2 (A). 

\section{Quantifying size and shape distributions}

 \begin{figure}
 \begin{center}
   \includegraphics[width=0.7\linewidth]{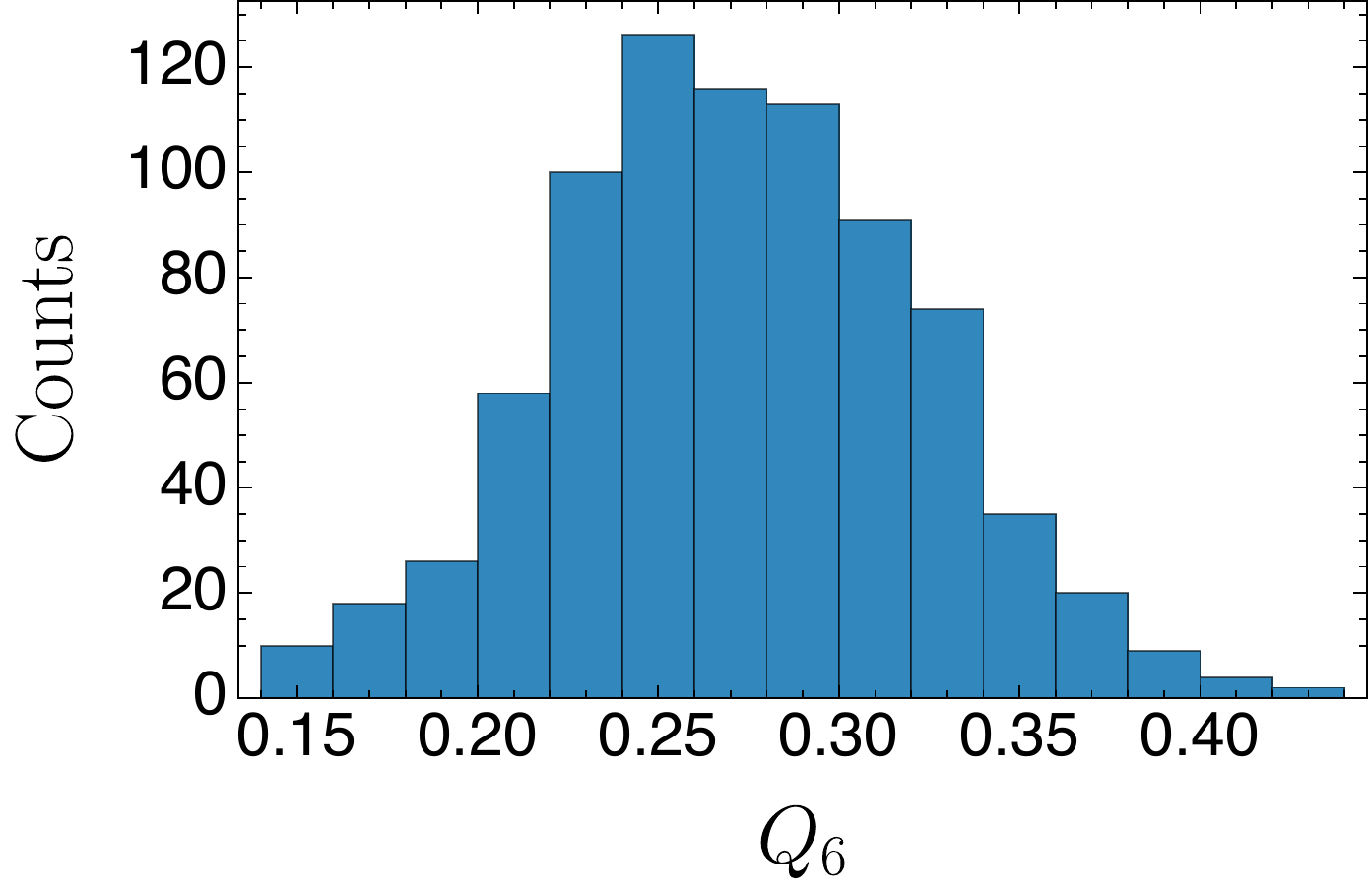}
   \includegraphics[width=0.7\linewidth]{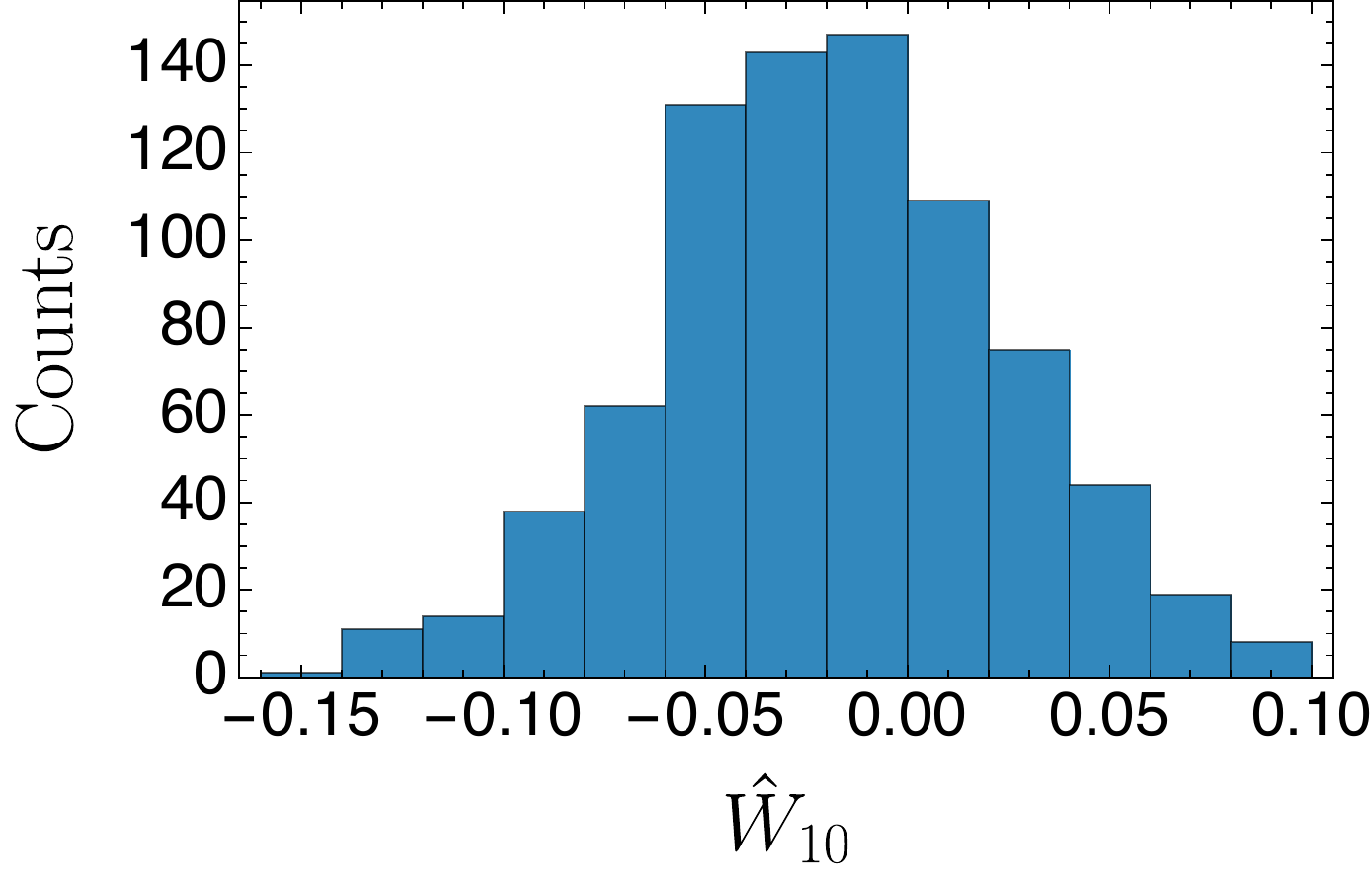}
 \end{center}
   \caption{Histograms of $Q_6$ and $\hat{W}_{10}$ from final structures of the simulations of the microscopic model.}
   \label{fig:bohists}
 \end{figure}

  \begin{figure}
 \begin{center}
   \includegraphics[width=0.7\linewidth]{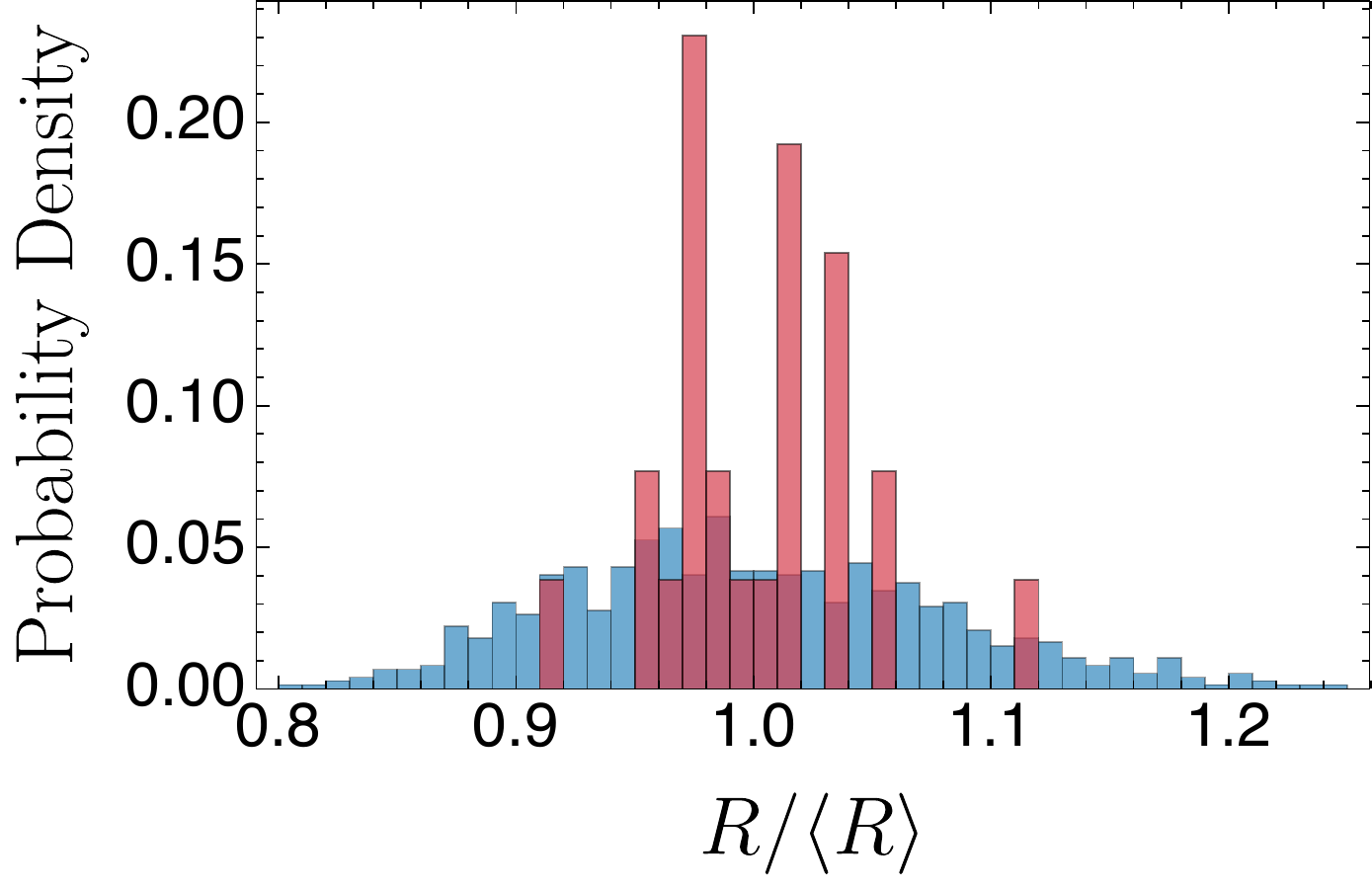}
    \end{center}
   \caption{Histograms of assembled microcompartments' effective
     diameter from simulation of our molecular model (blue). We also
     show experimental data (red) from Ref.~\cite{Iancu:2010eb}.}
   \label{fig:experi}
 \end{figure}

In order to quantify the extent of shape fluctuations, we computed both a measure of sphericity and a measure of the ideality of the icosahedral shape. 
We measure the sphericity by quantifying deviations from the mean radius.
Let $\mathbf{r}_0$ denote the center of mass of the vertices. 
We denote by $d_i$ magnitude of the vector $\mathbf{r}_i$ which connects $\mathbf{r}_0$ to a vertex indexed by $i$.
Then the sphericity is calculated as
\begin{equation}
 \alpha(\mathcal{V}) = 1 - \sum_{i\in \mathcal{V}} \left( \avg{d_i^2} - \avg{d_i}^2 \right) / \avg{d_i}.
 \end{equation}
 The number $\alpha$ is 1 on a perfect sphere.
 As the variance of the distance to shell grows, the value of the order parameter decreases.
 This parameter is not
 strongly influenced by the presence of sharp edges and vertices, but instead characterizes the extent to which a structure is isotropic rather than distended or elongated.

Bond orientational order parameters (cf. Ref.\cite{Steinhardt:1983uh}) provide a  very sensitive measure of icosahedral symmetry.
We computed the distribution of two such parameters that distinguish icosahedral orientational order from face-centered cubic, body-centered cubic, hexagonal close packed, and simple cubic packings. 
Following Steinhardt et al., we define
\begin{equation}
  Q_6(\{r_1,\dots, r_{N_\textrm{defect}}\}) = \frac{1}{N_\textrm{defect}} \sum_{i=1}^{N_\textrm{defect}} \sum_{l=-m}^m Y_{6m} (r_i)
\end{equation}
and 
\begin{equation}
\hat{W}_{10} = \sum_{m_1, m_2, m_3 = -l, \dots, l} \left( \begin{tabular}{ccc} $l$ & $l$ & $l$\\ $m_1$ & $m_2$ & $m_3$ \end{tabular}\right) \frac{Q_{l,m_1} Q_{l,m_2} Q_{l,m_3}}{\left( \sum_{m=-l}^l Q_{lm} \right)^{3/2} },
\end{equation}
where the coefficient is the Wigner 3-$j$ symbol.
For an ideal icosahedral structure
\begin{eqnarray}
  Q_6 \approx 0.66 & \hat{W}_{10} \approx -0.094
\end{eqnarray}
Previous studies have classified icosahedrality using a minimum of 0.6 for $Q_6$ or a maximum of $-0.1$ for $\hat{W}_{10}.$
By these metrics, our typical structures deviate strongly from perfect icosahedra. 
The distribution of $\hat{W}_{10}$ suggests that some structures
reasonably approximate
ideal icosahedra, but those structures are rare, in the tail of the distribution of Fig.~\ref{fig:bohists}.
In contrast, the $Q_6$ order parameter does not register icosahedrality in our structures, as seen in Fig.~\ref{fig:bohists}.

Though the structures are clearly not ideal icosahedra, they have the distinct faceted appearance of the structures typically seen in tomograms of carboxysomes and other bacterial microcompartments.
To quantify the extent to which the structure is faceted, it is useful to look at the extent of bending in the proximity of fivefold defects. 
We use the average bending angle of the five bonds around a defect as a faceting measure.
Let $\mathcal{V}_\textrm{d}$ denote the set fivefold defect vertices in a structure, of which there are $N_\textrm{d}.$ 
We also define $\mathcal{B}(v_i)$ to be the set of bonds connected to vertex $v_i$ which has elements $ij.$
Then, the faceting parameter is
\begin{equation}
  \avg{\theta_\textrm{defect}} = \frac{1}{5 N_\textrm{d}} \sum_{v_i\in \mathcal{V}_\textrm{d}} \sum_{ij \in \mathcal{B}(v_i)} | \theta_{ij} |.
\end{equation}

We quantified the size of assembled microcompartments according to their ``effective'' radii.
We defined the effective radius as
\begin{equation}
  R_\textrm{eff} = \left(\frac{3 N_{\textrm{c}}}{4\pi}\right)^{1/3},
\end{equation}
which assumes an idealized spherical droplet.  While this measure of
the radius is only approximate, it readily provides insight into the
degree of size heterogeneity.  A distribution of assembled compartment
size is shown in Fig.~\ref{fig:experi}, together with experimental data
reported in Ref.~\cite{Iancu:2010eb}.  It is
challenging to make a quantitative comparison in this case because the
number of experimental observations is small.

\clearpage

\end{document}